\documentclass[english]{IEEEtran}
\usepackage[T1]{fontenc}
\usepackage{color}
\usepackage{float}
\usepackage{amsmath}
\usepackage{amsthm}
\usepackage{amssymb}
\usepackage{stackrel}
\usepackage{graphicx}
\usepackage{multirow}
\usepackage{booktabs}
\usepackage{makecell}

\usepackage{caption}
\makeatletter

\floatstyle{ruled}
\newfloat{algorithm}{tbp}{loa}
\providecommand{\algorithmname}{Algorithm}
\floatname{algorithm}{\protect\algorithmname}

\theoremstyle{plain}
\newtheorem{prop}{\protect\propositionname}

\usepackage{amsfonts}
\usepackage{cite}
\usepackage{hyperref}   
\hypersetup{
    colorlinks=true,    
    linkcolor=blue,       
    citecolor=blue,    
    urlcolor=blue,      
    filecolor=blue        
}
\usepackage{array}
\usepackage{algorithm}
\usepackage{algorithmic}
\usepackage{subcaption}
 
\usepackage{stfloats}
\usepackage{graphicx}

\makeatother

\usepackage{babel}

\providecommand{\propositionname}{Proposition}

\begin{document}
\title{Predictive Control over Low-Altitude
Wireless Networks: Joint Trajectory Design and Resource Allocation}% with Finite Blocklength Transmission}
\author{ 
\IEEEauthorblockN{Haijia Jin},
\IEEEauthorblockN{Jun Wu},
\IEEEauthorblockN{Weijie Yuan, \IEEEmembership{Senior Member, IEEE}},
\IEEEauthorblockN{Ruizhi Ruan},
\IEEEauthorblockN{Jiacheng Wang, \IEEEmembership{Member, IEEE}},\\
\IEEEauthorblockN{Dusit Niyato, \IEEEmembership{Fellow, IEEE}},
\IEEEauthorblockN{Dong In Kim, \IEEEmembership{Life Fellow, IEEE}}, and
\IEEEauthorblockN{Abbas Jamalipour, \IEEEmembership{Fellow, IEEE}}
% \IEEEauthorblockN{Fan Liu, \IEEEmembership{Senior Member, IEEE}}, \\ and \IEEEauthorblockN{Yuanhao Cui, \IEEEmembership{Member, IEEE}}

\thanks{Haijia Jin, Jun Wu, Ruizhi Ruan, and Weijie Yuan are with the School of Automation and Intelligent Manufacturing, Southern University of Science and Technology, Shenzhen 518055, China (e-mails: \{jinhj2024, wuj2021, ruanrz2024\}@mail.sustech.edu.cn; yuanwj@sustech.edu.cn).

Jiacheng Wang and Dusit Niyato are with the College of Computing and Data Science, Nanyang Technological University, Singapore 639798 (e-mails: jiacheng.wang@ntu.edu.sg; dniyato@ntu.edu.sg).

Dong In Kim is with the Department of Electrical and Computer Engineering, Sungkyunkwan University, Suwon 16419, South Korea (e-mail: dongin@skku.edu).

Abbas Jamalipour is with the School of Electrical and Computer Engineering, University of Sydney, Australia, and with the Graduate School of Information Sciences, Tohoku University, Japan (e-mail: a.jamalipour@ieee.org).%

% \noindent\hangindent=1.5em\hangafter=1%  
%          \noindent\textbullet~ F. Liu is with the National Mobile Communications Research Laboratory,
%          Southeast University, Nanjing 210096, China (e-mail: f.liu@ieee.org).%
         
%\noindent\hangindent=1.5em\hangafter=1%  
 %        \noindent\textbullet~ Y. Cui is with the School of Information
 %and Communication Engineering, Beijing University of Posts and Telecom
%munications, Beijing 100876, China (e-mail: cuiyuan hao@bupt.edu.cn).

}
}

\IEEEtitleabstractindextext{
\begin{abstract}
 Low-altitude wireless networks (LAWNs) have been envisioned as flexible and transformative platforms for enabling delay-sensitive control applications in Internet of Things (IoT) systems. In this work, we investigate the real-time wireless control over LAWNs, where an aerial drone is employed to serve multiple mobile automated guided vehicles (AGVs) via finite blocklength (FBL) transmission. Toward this end, we adopt the model predictive control (MPC) to ensure accurate trajectory tracking, while we analyze the communication reliability using the outage probability. Subsequently, we formulate an optimization problem to jointly determine control policy, transmit power allocation, and drone trajectory by accounting for the maximum travel distance and control input constraints. To address the resultant non-convex optimization problem, we first derive the closed-form expression of the outage probability under FBL transmission. Based on this, we reformulate the original problem as a quadratic programming (QP) problem, followed by developing an alternating optimization (AO) framework. Specifically, we employ the projected gradient descent (PGD) method and the successive convex approximation (SCA) technique to achieve computationally efficient sub-optimal solutions. Furthermore, we thoroughly analyze the convergence and computational complexity of the proposed algorithm. Extensive simulations and AirSim-based experiments are conducted to validate the superiority of our proposed approach compared to the baseline schemes in terms of control performance.
\end{abstract}

}
\maketitle
\IEEEdisplaynontitleabstractindextext

\section{Introduction}
\subsection{Background}
The rapid evolution of the Internet of Things (IoT) has driven the deployment of large-scale, distributed systems that require robust, low-latency, and adaptive wireless connectivity. To meet these demands, low-altitude wireless networks (LAWNs) have emerged as a promising paradigm for near-ground IoT applications such as smart factories, urban mobility, and intelligent logistics \cite{LAWN_IoT}. Thanks to their close proximity to ground-level devices, LAWN architectures offer several inherent advantages over conventional terrestrial or satellite networks, including flexible deployment, enhanced spatial resolution, and the ability to maintain line-of-sight (LoS) links in complex or dynamic environments. Furthermore, the low-altitude operational domain enables LAWNs to accommodate dense device connectivity and facilitate swift topological adjustments, rendering them ideal for time-critical and mission-sensitive IoT services \cite{LAWN2,wu2025low,yuan}.

Within LAWNs, drone-assisted communication has emerged as a cornerstone technology for enabling reliable and flexible wireless connectivity in dynamic, large-scale, and infrastructure-limited environments. Drones can exploit their aerial mobility and LoS advantages to dynamically establish high-quality links with ground nodes, thereby supporting low-latency and resilient data transmission \cite{Wang_Dusit,UAV_IoT1,UAV_IoT2,URA}. To fully harness these benefits, extensive studies have been conducted on optimizing drone deployment, user association, and spectrum allocation. For example, Mozaffari \emph{et al.} \cite{UAV_communication} formulated a joint optimization framework for drone mobility, device association, and power control in IoT networks. In addition, adaptive drone trajectory designs have been proposed to improve link quality and enhance security, particularly in mobile or adversarial environments \cite{UAV_com_moving2,DI_KIM}. These efforts have significantly advanced the capabilities of drone-enabled communication services in LAWN scenarios.

Complementary to the communication layer, real-time control of ground agents like automated guided vehicles (AGVs) constitutes another critical function within LAWN applications, particularly in domains such as industrial automation, logistics, and smart transportation \cite{IoT_AGV1,AGV_tracking1}. Among various control strategies, model predictive control (MPC) stands out for its ability to anticipate future trajectories and optimize control actions with physical and operational constraints in a receding-horizon manner. Compared with classical approaches such as proportional-integral-derivative (PID) and linear quadratic regulator (LQR) controllers \cite{MPC_LQR,PID}, MPC offers greater adaptability and performance robustness in complex and dynamic environments. It has been widely applied in AGV trajectory tracking, multi-agent coordination, and collision avoidance tasks \cite{MPC1,MPC2,MPC3}. However, in most existing studies, communication and control are treated as two independent subsystems. Communication is responsible for ensuring data delivery, while control focuses on generating optimal motion commands, without explicitly modeling their mutual dependency, leading to suboptimal system behavior, especially when time delays, packet losses, or limited bandwidth affect control execution.

To address the growing need for real-time responsiveness and reliability in LAWNs, recent research has increasingly focused on the joint optimization of communication and control, giving rise to the concept of wireless networked control systems (WNCSs). Unlike conventional approaches that design communication protocols and control strategies separately, WNCS aims to coordinate both subsystems in a unified manner to enhance closed-loop performance under stringent bandwidth, latency, and energy constraints \cite{WNCS1,WNCS2,SC2,WNCS3}. The integrated perspective stems from the recognition that control performance is fundamentally influenced by the quality and timeliness of information exchange, particularly in time-sensitive applications such as trajectory tracking and multi-agent coordination.
While existing joint design studies focus on trajectory planning and resource allocation at the \textit{outer loop} level, they typically overlook how communication unreliability affects the stability of \textit {inner loop} control mechanisms, such as drone attitude regulation and real-time feedback control in LAWNs \cite{control_difference}.
Toward this end, Kostina \emph{et al.} \cite{rate_cost_trade_off} established a fundamental lower bound on the minimum communication rate required to achieve a specified LQR cost, providing an information-theoretic benchmark for control-aware communication design. Building on this, subsequent works have formulated cross-layer optimization problems that jointly consider control accuracy and communication throughput, as exemplified by recent efforts in integrated sensing-communication-computing-control ($\mathbf{SC^3}$) architectures for satellite-drone systems \cite{cont&comm4}. Meanwhile, algorithmic advances have sought to address practical challenges in wireless environments. For instance, Cao \emph{et al.} \cite{WNCS_MPC2} developed a robust MPC framework that dynamically adapts to variable delays, measurement noise, and packet drops, thus ensuring reliable AGV tracking even under fluctuating network conditions. Despite these developments, most existing approaches implicitly assume the availability of large-packet or infinite blocklength (IBL) transmissions, which do not accurately reflect the stringent latency and reliability requirements encountered in real-time control systems. In such systems, operational periods are typically segmented into numerous short time slots to support agile communication and timely control decisions \cite{real_time}, making short-packet transmission not only common but also essential.

To address this mismatch, finite blocklength (FBL) transmission has been adopted as a more realistic communication model in real-time control settings. Compared to IBL transmission, where the achievable rate follows Shannon capacity, FBL introduces a more complex rate expression that depends on the signal-to-noise ratio (SNR), block error rate (BLER), and blocklength \cite{FBL_Rate}, which poses significant challenges for system optimization, as the rate function under FBL is typically non-convex and highly sensitive to system parameters, thereby making resource allocation and control-communication co-design considerably more difficult than in IBL-based systems \cite{FBL11}. Recent studies have applied FBL theory in various drone communication scenarios. For instance, the secrecy and covertness performance of FBL-based drone communication was investigated in \cite{UAV_FBL2} and \cite{UAV_FBL3}, while Ruat \emph{et al.} \cite{UAV_FBL4} analyzed reliability optimization in a drone-assisted nonlinear energy harvesting full-duplex network under both IBL and FBL regimes.

\subsection{Motivation and Contributions}
\textcolor{black}{Despite the aforementioned efforts and the substantial progress achieved in LAWN-enabled control systems, to the best of our knowledge, most existing studies have primarily focused on either optimizing control strategies under communication constraints or enhancing wireless resource allocation to support control performance, while often overlooking the intricate interplay between communication reliability and control actions, particularly in latency-critical real-time control scenarios \cite{WNCS1,WNCS2,SC2,WNCS3,cont&comm4,WNCS_MPC2}. In addition, many works restrict attention to single-agent settings instead of addressing coordinated control of multiple AGVs \cite{MPC1,MPC2,MPC3,WNCS_MPC2}. Motivated by the challenges of real-time control and communication in dynamic environments, this paper develops a communication-assisted predictive control framework over LAWN. Specifically, we explore a real-time control scenario where an aerial drone serves multiple mobile AGVs via FBL transmissions. To achieve accurate and delay-sensitive trajectory tracking, we employ an MPC strategy, jointly optimizing the control policy, transmit power, and drone trajectory. Our goal is to minimize the control cost while explicitly accounting for communication reliability in the control formulation, subject to physical constraints such as mobility and power limits. The main contributions of this work are summarized as follows:
}

\textcolor{black}{1) We develop a novel wireless control framework for the drone-assisted LAWN system, enabling real-time remote trajectory tracking of multiple AGVs under FBL transmission. We use the MPC method to guide the drone's real-time decision-making, while analyzing communication reliability through outage probability to ensure reliable wireless transmission for precise and low-latency trajectory tracking.}

\textcolor{black}{2) To accomplish the control task, we formulate a joint optimization problem that minimizes the total control cost of real-time AGV operations by optimizing control inputs, power allocation, and drone trajectory, subject to constraints on control inputs and communication quality-of-service (QoS). We improve the framework further by incorporating the outage probability into the control cost function, explicitly capturing the impact of communication disruptions on the control performance with additional constraints on transmit power, actuation limits, and drone mobility, which are introduced to ensure the practical feasibility of the system design.}

%% unrevised
% Motivated by the above, this paper investigates the interaction between predictive control and wireless communication in the context of LAWN. Specifically, we consider a real-time control scenario where an aerial drone serves multiple mobile AGVs via FBL transmissions. To enable accurate and delay-sensitive trajectory tracking, we adopt an MPC strategy and jointly optimize the control policy, transmit power, and drone trajectory. Under this framework, our goal is to minimize the control cost while ensuring communication reliability and satisfying physical constraints such as mobility and power limits. The main contributions of this work are summarized as follows.

% 1) We develop a joint communication-control framework for the drone-assisted LAWN system, enabling real-time remote control of multiple AGVs under FBL transmission. We adopt the MPC method to guide the drone’s real-time decision-making. Moreover, we analyze the communication reliability using the outage probability to ensure reliable wireless transmission and support accurate, low-latency trajectory tracking.

% 2) To accomplish the control task, we formulate a joint optimization problem to minimize the total control cost of real-time AGV operations by jointly optimizing control inputs, power allocation, and drone trajectory, subject to constraints on maximum transmit power, actuation limits, and drone mobility and flight region.

3) To address the non-convex optimization problem, we first derive the closed-form expression of the outage probability under FBL transmission. Based on this, we recast the formulated non-convex problem into a quadratic programming (QP) structure and decompose it into a sequence of tractable sub-problems. We then develop an alternating optimization (AO) framework to iteratively update control actions, power allocation, and drone trajectory by leveraging projected gradient descent (PGD) and successive convex approximation (SCA). We rigorously analyze the convergence properties and computational complexity of the proposed algorithm to validate its practicality and scalability.

4) Extensive simulations confirm that the proposed algorithm enables the drone to robustly track AGVs even under abrupt trajectory variations and consistently outperforms existing benchmarks in control performance. Its effectiveness is further validated on the AirSim platform \cite{AirSim}, offering practical insights into the design of drone-assisted wireless control systems in LAWN.

The remainder of this article is organized as follows. Section \ref{sec:2} introduces the drone-assisted LAWN control network. Section \ref{sec:Problem-formulation} formulates the joint optimization of the MPC-based control input, power allocation, and drone trajectory, which is subsequently addressed in Section \ref{sec:Proposed-solutions}. Simulation and experimental results are presented in Section \ref{sec:4}, followed by the conclusions in Section \ref{sec:5}.

\textit{Notations: } Unless otherwise specified, bold lowercase and uppercase letters, e.g., $\mathbf{m}$ and $\mathbf{M}$, represent vectors and matrices, respectively. $\mathbb{R}^{M}$ denotes the space of $M$-dimensional real-valued column vectors. $\mathbb{E}\left(\cdot\right)$, $\|\cdot\|$, and $(\cdot)^T$ denote the expectation, $\ell_{2}$ norm, and matrix transposition, respectively. $\mathcal{N}(\mu,\sigma^{2})$ denotes the Gaussian distribution with mean $\mu$ and variance $\sigma^{2}$.  $\mathrm{Pr}(\cdot)$ represents the probability of an event. In addition, $\mathbf{I}_{M}$ denotes the $M \times M$ identity matrix, and the function $Q^{-1}(\cdot)$ is the inverse of the Gaussian $\mathcal{Q}$-function, defined as $Q(x)=\frac{1}{\sqrt{2\pi}}\int_{x}^{\infty}\exp(-\frac{t^{2}}{2})\textrm{d}t$. Moreover, to facilitate understanding of the mathematical derivations, the key symbols used throughout the system and optimization models are summarized in Table~\ref{tab:notations}.

\begin{table*}[!t]
\centering
\caption{List of key symbols.}
\label{tab:notations}
\begin{tabular}{clcl}
\toprule
\textbf{Symbol} & \textbf{Description} & \textbf{Symbol} & \textbf{Description} \\
\midrule
$\mathbf{A}_{k}$, $\mathbf{B}_{k}$ & State transition and input matrices &
$\tilde{\mathbf{A}}_{k}$, $\tilde{\mathbf{B}}_{k}$ & Augmented system matrices \\

$B$ & Channel bandwidth &
$\mathbf{c}_{k}[n]$ & Horizontal position of the $k$-th AGV \\

$\mathcal{D}$ & Aerial drone flight region &
$d_k[n]$ & Distance between drone and AGV \\

$\mathbf{e}_{k}[n]$ & Predicted deviation &
$f_{\mathrm{QoS}}$ & Communication QoS \\

\makecell[l]{$\mathbf{G}_{k}[n]$, $\mathbf{H}_{k}[n]$, $M_{k}[n]$,\\ $L_{k}[n]$, $S_{k}[n]$, $W_{k}[n]$} & Coefficients of reformulated QP &
$H$ & Fixed altitude of aerial drone \\

$h_k[n]$ & Channel between drone and AGV &
$\hat{h}_k[n]$ & Channel small-scale fading \\

$J_n$ & Control cost function &
$\hat{J}_{n}$ & Control cost of reformulated QP \\

$K$ & Number of AGVs &
$\mathcal{K}$ & Set of AGVs \\

$l_k$ & Blocklength of transmitted signal &
$m$ & Nakagami shape parameter \\

$N$ & Number of discrete time slots &
$N_p$ & Prediction horizon of MPC \\

$P_k[n]$ & Outage probability &
$P_{\max}$ & Maximum transmit power \\

$p_k[n]$ & Transmit power &
$\mathbf{q}[n]$ & Horizontal coordinate of aerial drone \\

$\mathbf{Q}_1,\mathbf{Q}_2,\mathbf{Q}_3$ & Weighting matrices of control cost &
$\mathbf{Q}_4,\mathbf{Q}_5$ & Weighting matrices of reformulated QP \\

$R_k[n]$, $R_k^{\text{th}}$ & Achievable rate and rate threshold &
$T$ & Total time duration \\

$\boldsymbol{\mathcal{U}}_{k}$ & Predefined control set &
$\mathbf{u}_{k}[n]$ & Control input \\

$V_k[n]$ & Channel dispersion &
$V_{\max}$ & Maximum drone speed \\

$\mathbf{x}_{k}[n]$ & AGV's state vector &
$\tilde{\mathbf{x}}_{k}[n]$ & Augmented AGV state \\

$\tilde{\mathbf{x}}_{k,\textrm{ref}}[n]$ & Reference AGV state &
$\alpha_0$ & Reference channel gain at 1 m \\

$\Gamma_k[n]$, $\Gamma_k^{\text{th}}$ & SNR and SNR threshold &
$\Delta \tau$ & Duration of each time slot \\

$\Delta\mathbf{u}_{k}[n]$ & Augmented control input &
$\Delta\boldsymbol{\mu}_{k}[n]$ & Control input sequence over horizon \\

$\epsilon$ & BLER &
$\sigma_k^2$ & Noise power \\

$\boldsymbol{\chi}_{k}[n]$ & Predicted control state vector &
$\boldsymbol{\chi}_{k,\textrm{ref}}[n]$ & Reference state over horizon \\
\bottomrule
\end{tabular}
\end{table*}

\begin{center}
\begin{figure}
\begin{centering}
\includegraphics[width=1\linewidth]{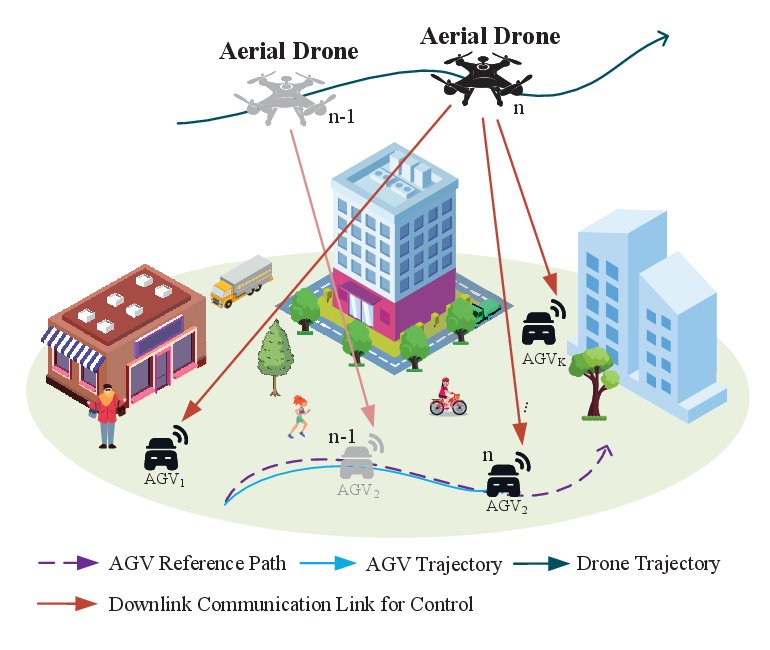}
\par\end{centering}
\centering{}
\captionsetup{justification=raggedright,singlelinecheck=false} % 设置标题靠左对齐
\caption{\label{fig:sys_mod}The considered LAWN comprises a mobile drone enabling wireless control for AGV path tracking.}
\end{figure}
\par\end{center}

\section{System Model \label{sec:2}}

As shown in Fig. \ref{fig:sys_mod}, we investigate a LAWN system, in which a single-antenna drone remotely controls $K$ AGVs for real-time trajectory tracking. At each time slot $n$, the horizontal coordinate of the drone is denoted as $\mathbf{q}\left[n\right]=\left[x\left[n\right],y\left[n\right]\right]^{\textrm{T}}$ with a constant altitude $H$.\footnote{\textcolor{black}{To avoid collisions with ground infrastructure and meet airspace regulations, the drone is set to fly at a fixed altitude, while ensuring stable communication coverage \cite{SC2}.}} 
The $K$ AGVs are indexed by $\ensuremath{\mathcal{K}=\left\{ 1,2,\ldots, K\right\} }$, and the location of $k$-th AGV is represented as $\mathbf{c}_{k}\left[n\right]=\left[x_{k}\left[n\right],y_{k}\left[n\right]\right]^{\textrm{T}},\forall k\in\mathcal{K}$, assuming zero altitude. To ensure reliable and dynamic trajectory tracking, the drone, functioning as a remote controller, generates optimal control actions based on the AGVs’ current states and transmits the corresponding commands to the AGVs over the wireless channel \cite{WNCS_MPC2}.\footnote{\textcolor{black}{In this study, we assume ideal uplink channels to enable perfect state estimation of all AGVs, focusing on the impact of downlink communication on control performance, which has also been considered in prior studies such as \cite{cont&comm4,MPC_increment}.}} To facilitate system design, we consider a total duration $T$ divided into $N$ discrete time slots, each with a duration $\Delta \tau = T/N$. Moreover, $\Delta \tau$ is chosen to be sufficiently small to ensure that the states of the AGVs remain approximately constant during each time slot.

\subsection{Communication Model}

At each time slot, the drone transmits the control commands to the AGVs over the frequency orthogonal channel. Denote the real-time distance between the drone and the $k$-th AGV as
\begin{equation}
d_{k}\left[n\right]=\sqrt{\parallel\mathbf{q}\left[n\right]-\mathbf{c}_{k}\left[n\right]\parallel^{2}+H^{2}}, \forall k\in\mathcal{K}.
\end{equation}
\textcolor{black}{The realistic air-to-ground (A2G) channel can be expressed as
\begin{equation}
h_{k}[n] = \sqrt{\frac{\alpha_{0}}{d_{k}^{2}[n]}} \, \hat{h}_{k}[n], \forall k\in\mathcal{K},
\end{equation}
where $\alpha_{0}$ is the channel gain at the reference distance of $d_{k}[n] = 1\,\mathrm{m}$, and $\alpha_{0}/d_{k}^{2}[n]$ characterizes the large-scale path loss.
% \footnote{\textcolor{black}{In practical LAWN deployments, perfect channel state information (CSI) is difficult to obtain, which motivates robust CSI-aware designs for real-world systems. However, since the primary objective of this work is to investigate the impact of communication reliability on control performance, we follow the common assumption of perfect CSI, as in \cite{10279816,9557830,9724198}, and leave robustness to CSI uncertainty for future work.}}
Moreover, the small-scale fading component $\hat{h}_k[n]$ is modeled as Nakagami-$m$ distributed, which has been widely adopted to characterize low-altitude A2G channels in LAWN deployments, effectively capturing both LoS and non-line-of-sight (NLoS) propagation \cite{nakagami1,nakagami2}.
% Moreover, the small-scale fading component, $\hat{h}_k\left[n\right]$, follows a Nakagami-$m$ distribution, which effectively captures both LoS and non-line-of-sight (NLoS) propagation, as widely reported in the literature, e.g., \cite{nakagami1}, \cite{nakagami2}. 
The fading parameter $m$ quantifies the severity of fading, where a larger $m$ corresponds to weaker fading and a higher probability of LoS dominance. 
% \footnote{\textcolor{black}{Prior studies \cite{nakagami3}, \cite{nakagami4} have demonstrated that CE errors affect the fading parameter of the estimated channel, resulting in an adjusted parameter $\tilde{m}$. As the average SNR or the number of pilot symbols increases, $\tilde{m}$ converges to $m$. For analytical simplicity, we assume $\tilde{m}=m$, implying that the estimated channel $\tilde{h}_{k}[n]$ also follows a Nakagami-$m$ distribution with $\mathbb{E}\left[\lvert\tilde{h}_{k}[n]\lvert^2\right]=\alpha_{0}/d_{k}^{2}[n]$.}}
Accordingly, the instantaneous SNR at the $k$-th AGV is given by
 \begin{equation}
    \Gamma_{k}\left[n\right]=\frac{\lvert h_{k}\left[n\right]\lvert^{2}p_{k}\left[n\right]}{\sigma_{k}^{2}}, \forall k \in \mathcal{K},\label{SINR}
\end{equation}
where $p_{k}\left[n\right]$ denotes the transmit power allocated to the $k$-th AGV and $\sigma_{k}^{2}$ represents the noise power at the receiver.}
The achievable communication rate from the drone to the $k$-th AGV is $R_{k}\left[n\right]=B\log_{2}\left(1+\Gamma_{k}\left[n\right]\right)$ with $B$ denoting the sub-channel bandwidth. In order to achieve delay-critical control tasks in LAWNs, this work adopts FBL transmission to meet stringent latency requirements. The achievable communication rate under FBL transmission depends on the SNR, the BLER, and the blocklength, given by \cite{FBL_Rate} 
\begin{equation}
R_{k}\left[n\right]=B\left(\log_{2}\left(1+\Gamma_{k}\left[n\right]\right)-\sqrt{\frac{V_{k}\left[n\right]}{l_{k}}}Q^{-1}\left(\epsilon\right)\right), \forall k \in\mathcal{K},  
\end{equation}
where $l_{k}$ denotes the blocklength of the transmitted signal, $\epsilon$ represents the BLER, and $V_{k}\left[n\right]$ is the channel dispersion, which is defined as
\begin{equation}
V_{k}\left[n\right]=1-\left(1+\Gamma_{k}\left[n\right]\right)^{-2}, \forall k\in\mathcal{K}.
\end{equation}

\subsection{Control Model}

As for the control part, we model the aerial drone and each AGV as a linear control system. The discrete-time dynamics of the $k$-th control system are described by\footnote{For analytical tractability, we neglect the control system noise at the AGVs in this study \cite{WNCS_MPC}.}
\begin{equation}
\mathbf{x}_{k}\left[n+1\right]=\mathbf{A}_{k}\mathbf{x}_{k}\left[n\right]+\mathbf{B}_{k}\mathbf{u}_{k}\left[n\right], \label{control_system_equation}
\end{equation}
\textcolor{black}{where $\mathbf{x}_{k}\left[n\right]=\left[x_{k}\left[n\right],y_{k}\left[n\right],v_{x,k}\left[n\right],v_{y,k}\left[n\right]\right]^{\mathrm{T}}\in\mathbb{R}^{4}$ denotes the system state of the $k$-th AGV with $\left[v_{x,k}\left[n\right],v_{y,k}\left[n\right]\right]$ representing the AGV's velocity during the time interval $\left(n\Delta \tau,\left(n+1\right)\Delta \tau\right]$. The control input $\mathbf{u}_{k}\left[n\right]=\left[a_{x,k}\left[n\right],a_{y,k}\left[n\right]\right]^{\mathrm{T}}\in\mathbb{R}^{2}$ denotes the commanded acceleration vector at time slot $n$. The matrices $\mathbf{A}_{k}\in\mathbb{R}^{4\times4}$ and $\mathbf{B}_{k}\in\mathbb{R}^{4\times2}$ represent the discrete-time state transition and input matrices under the constant-acceleration model, respectively, given by
\begin{equation}
    \mathbf{A}_{k}=\left[\begin{array}{cccc}
1 & 0 & \Delta\tau & 0\\
0 & 1 & 0 & \Delta\tau\\
0 & 0 & 1 & 0\\
0 & 0 & 0 & 1
\end{array}\right],\mathbf{B}_{k}=\left[\begin{array}{cc}
\frac{\Delta\tau^{2}}{2} & 0\\
0 & \frac{\Delta\tau^{2}}{2}\\
\Delta\tau\  & 0\\
0 & \Delta\tau
\end{array}\right].
\end{equation}
To ensure smooth actuation and avoid unrealistic acceleration jumps between consecutive sampling steps, the original system in \eqref{control_system_equation} is reformulated by introducing the acceleration increment $\Delta\mathbf{u}_{k}\left[n\right]=\mathbf{u}_{k}\left[n\right]-\mathbf{u}_{k}\left[n-1\right]$ as the new control input \cite{MPC_increment}. Correspondingly, the current state $\mathbf{x}_{k}\left[n\right]$ together with the previous control input $\mathbf{u}_{k}\left[n-1\right]$ are stacked into an augmented state vector, i.e.,}
\begin{equation}
   \tilde{\mathbf{x}}_{k}\left[n\right]=\left[\begin{array}{c}
\mathbf{x}_{k}\left[n\right]\\
\mathbf{u}_{k}\left[n-1\right]
\end{array}\right].
\end{equation}
The resulting augmented discrete-time system evolves as
\begin{equation}
\tilde{\mathbf{x}}_{k}\left[n+1\right]=\mathbf{\tilde{A}}_{k}\tilde{\mathbf{x}}_{k}\left[n\right]+\mathbf{\tilde{B}}_{k} \Delta\mathbf{u}_{k}\left[n\right],\label{system_equation_new}
\end{equation}
where $\mathbf{\tilde{A}}_{k}\in\mathbb{R}^{6\times6}$ and $\mathbf{\tilde{B}}_{k}\in\mathbb{R}^{6\times2}$ represent the updated state transition and input matrices, respectively, and are defined as
\begin{equation}
\mathbf{\tilde{A}}_{k}=\left[\begin{array}{cc}
\mathbf{A}_{k} & \mathbf{B}_{k}\\
\mathbf{0}_{2\times4} & \mathbf{I}_{2\times2}
\end{array}\right],\mathbf{\tilde{B}}_{k}=\left[\begin{array}{c}
\mathbf{B}_{k}\\
\mathbf{I}_{2\times2}
\end{array}\right].
\end{equation}

\section{Problem formulation \label{sec:Problem-formulation}}
 In this section, we first present a general MPC framework that aims to minimize the control cost over a finite horizon subject to communication-induced constraints, followed by extending this formulation to account for the impact of transmission outage on control effectiveness.
\subsection{General Framework}
To ensure precise trajectory tracking under dynamic conditions, this work adopts an MPC framework. The goal of the MPC framework is to generate an optimal sequence of control inputs that minimizes the deviation between predicted system states and reference trajectories within a finite prediction horizon of length $N_{p}$. The associated control cost function is given by \eqref{control_cost} \cite{MPC_increment},
\begin{figure*}[tp]
\begin{equation}
J_{n}=\sum_{k\in\mathcal{K}}\sum_{i=0}^{N_{p}-1}\mathbf{e}^{\textrm{T}}_{k}\left[n+i \lvert n\right]\mathbf{Q}_{1}\mathbf{e}_{k}\left[n+i \lvert n\right]+ \Delta\mathbf{u}^{\textrm{T}}_{k}\left[n+i \lvert n\right]\mathbf{Q}_{2} \Delta\mathbf{u}_{k}\left[n+i \lvert n\right]+\mathbf{e}^{\textrm{T}}_{k}\left[n+N_{p}\lvert n\right]\mathbf{Q}_{3}\mathbf{e}_{k}\left[n+N_{p}\lvert n\right].\label{control_cost}
\end{equation}
\hrule
\end{figure*}
as shown at the top of the next page, where 
\begin{equation}
    \mathbf{e}_{k}\left[n+i\lvert n\right]=\mathbf{\tilde{\mathbf{x}}}_{k}\left[n+i\lvert n\right]-\mathbf{\tilde{\mathbf{x}}}_{k,\textrm{ref}}\left[n+i\right],
\end{equation}
denotes the predicted deviation from the reference state for the $k$-th AGV at time step $n+i$ and $\Delta\mathbf{u}_{k}\left[n+i\lvert n\right]$ represents the predicted control input. 
\textcolor{black}{The first term in \eqref{control_cost} penalizes the tracking error to ensure that the AGV trajectories remain close to their desired paths, while the second term penalizes large control variations to guarantee smooth actuation and prevent excessive acceleration. The positive semi-definite weighting matrices $\mathbf{Q}_{1}\in\mathbb{R}^{6\times6}$ and $\mathbf{Q}_{2}\in\mathbb{R}^{2\times2}$ determine the relative importance between tracking precision and control effort. The third term serves as a terminal cost to promote long-term control stability and reduce steady-state error. The matrix $\mathbf{Q}_{3}\in\mathbb{R}^{6\times6}$ is obtained by solving the discrete Riccati equation \cite{MPC1}
\begin{equation}
   \begin{split}\mathbf{Q}_{3}= &\  \mathbf{Q}_{1}+\mathbf{\tilde{A}}_{k}^{\textrm{T}}\mathbf{Q}_{3}\mathbf{\tilde{A}}_{k}-\mathbf{\mathbf{\tilde{A}}}_{k}^{\textrm{T}}\mathbf{Q}_{3}\mathbf{\tilde{B}}_{k}\\
 & \times \left(\mathbf{Q}_{2}+\mathbf{\tilde{B}}_{k}^{\textrm{T}}\mathbf{Q}_{3}\mathbf{\tilde{B}}_{k}\right)^{-1}\mathbf{\mathbf{\tilde{B}}}_{k}^{\textrm{T}}\mathbf{Q}_{3}\mathbf{\tilde{A}}_{k},
\end{split} 
\end{equation}
which balances immediate and future tracking performance within the finite prediction horizon.}
% ///original
% $\mathbf{Q}_{1}\in\mathbb{R}^{6\times6}$ and $\mathbf{Q}_{2}\in\mathbb{R}^{2\times2}$ are positive semi-definite weighting matrices used to balance the trade-off between tracking accuracy and control energy consumption. The term $\mathbf{e}^{\textrm{T}}_{k}\left[n+N_{p}\lvert n\right]\mathbf{Q}_{3}\mathbf{e}_{k}\left[n+N_{p}\lvert n\right]$ is introduced to enhance long-term control performance, with $\mathbf{Q}_{3}\in\mathbb{R}^{6\times6}$ being the solution of a discrete Riccati equation \cite{MPC1}
% \begin{equation}
%    \begin{split}\mathbf{Q}_{3}= &\  \mathbf{Q}_{1}+\mathbf{\tilde{A}}_{k}^{\textrm{T}}\mathbf{Q}_{3}\mathbf{\tilde{A}}_{k}-\mathbf{\mathbf{\tilde{A}}}_{k}^{\textrm{T}}\mathbf{Q}_{3}\mathbf{\tilde{B}}_{k}\\
%  & \times \left(\mathbf{Q}_{2}+\mathbf{\tilde{B}}_{k}^{\textrm{T}}\mathbf{Q}_{3}\mathbf{\tilde{B}}_{k}\right)^{-1}\mathbf{\mathbf{\tilde{B}}}_{k}^{\textrm{T}}\mathbf{Q}_{3}\mathbf{\tilde{A}}_{k}.
% \end{split} 
% \end{equation}
To facilitate system design, all predictive control input increments for the $k$-th AGV over the prediction horizon are organized into a vector $\Delta\boldsymbol{\mu}_{k}\left[n\right]\in\mathbb{R}^{2N_{p}}$, which compactly represents its acceleration decisions at each future time step, defined as 
\begin{equation}
\Delta\boldsymbol{\mu}_{k}\left[n\right]=\left[\begin{array}{c}
\Delta\mathbf{u}_{k}\left[n\lvert n\right]\\
\Delta\mathbf{u}_{k}\left[n+1\lvert n\right]\\
\vdots\\
\Delta\mathbf{u}_{k}\left[n+N_{p}-1\lvert n\right]
\end{array}\right] .
\end{equation}
\textcolor{black}{In the considered scenario, control commands are transmitted over wireless links, and the reliability of these links directly affects control performance. To account for this, the control input optimization incorporates a communication QoS constraint, ensuring that the control actions are optimized under communication conditions that guarantee successful transmission. This formulation explicitly couples control performance with communication reliability, ensuring that the control policy is computed only when reliable communication conditions are met, which will be detailed in the next subsection. The general optimization problem is thus expressed as}
% ///original
% In the considered scenario, control commands are transmitted over wireless links, and thus, communication reliability has a significant impact on overall control performance. To compromise, the control input optimization must explicitly incorporate communication QoS constraints, such that the general form of the optimization problem can be formulated as
\begin{subequations}\label{P0}
\begin{align}
\min_{\left\{\Delta\boldsymbol{\mu}_{k}[n]\right\},\mathbf{p}[n],\mathbf{q}[n]} \quad & J_n \label{p0}\\
\mathrm{s.t.} \quad & \Delta\boldsymbol{\mu}_k[n] \in \boldsymbol{\mathcal{U}}_{k},\forall k \in \mathcal{K}, \label{P0-C1}\\
& f_{\mathrm{QoS}}\in \mathcal{C}, \label{P0-C2}%
\end{align}%
\label{eq:abstract_joint_optimization}%
\end{subequations}
where $\mathbf{p}\left[n\right]=\left[p_{1}\left[n\right],\ldots,p_{K}\left[n\right]\right]^{\textrm{T}}$ denotes the power allocation vector, $\boldsymbol{\mathcal{U}}_{k}$ is the predefined control set corresponding to the $k$-th AGV, and $f_{\mathrm{QoS}}$ represents a communication quality metric that must meet $
\mathcal{C}$ to ensure the required QoS performance.

\subsection{Outage Probability-Based Formulation  }
To explicitly characterize how communication reliability influences control, we model the real-time transmission reliability in terms of outage probability. A communication outage is defined as when the real-time communication rate $R_{k}[n]$ falls below the QoS threshold, preventing the successful delivery of control commands to the AGVs. As a result, during the outage periods, the AGV system evolves based solely on its intrinsic physical dynamics in the absence of external control input. Accordingly, the outage probability of drone-to-AGV transmission is given by
\begin{equation}
    P_{k}\left[n\right]=\Pr\left(R_{k}\left[n\right]<R_{k}^{\textrm{th}}\right),\forall k \in\mathcal{K},\label{outage1}
\end{equation}
where $R_{k}^{\textrm{th}}$ denotes the minimum required data rate to satisfy the QoS constraint for the $k$-th AGV. To account for the influence of wireless communication, the discrete-time control system is accordingly reformulated as
\begin{equation}
\begin{split}\tilde{\mathbf{x}}_{k} & \left[n+1\right]\\
= & \begin{cases}
\mathbf{\tilde{A}}_{k}\tilde{\mathbf{x}}_{k}\left[n\right]+\mathbf{\tilde{B}}_{k} \Delta\mathbf{u}_{k}\left[n\right] & \textrm{with a prob. of }\left(1-P_{k}\left[n\right]\right),\\
\mathbf{\tilde{A}}_{k}\tilde{\mathbf{x}}_{k}\left[n\right] & \textrm{with a prob. of }P_{k}\left[n\right].
\end{cases}
\end{split}
\label{csowc}
\end{equation}
Accordingly, the expected system state of the $k$-the AGV at the next time instant is formulated as
\begin{equation}
\mathbb{E}\left[\tilde{\mathbf{x}}_{k}\left[n+1 \right]\right] =\mathbf{\tilde{A}}_{k}\tilde{\mathbf{x}}_{k}\left[n\right]+\left(1-P_{k}\left[n\right]\right)\mathbf{\mathbf{\tilde{B}}}_{k} \Delta\mathbf{u}_{k}\left[n \right].\label{expected_control_equation} 
\end{equation}
Substituting the expected dynamics into the original cost yields a modified objective that incorporates the impact of transmission unreliability. The resultant optimization problem is formulated as
\begin{subequations}\label{P1}
\begin{align}
\min_{\left\{\Delta\boldsymbol{\mu}_{k}\left[n\right]\right\},\mathbf{p}\left[n\right],\mathbf{q}\left[n\right]} & \quad \mathbb{E}\left(J_n\right)\\
\mathrm{s.t.} & \quad \sum_{k\in\mathcal{K}}p_{k}\left[n\right]\leq P_{\max},\label{P1-C1}\\
& \quad p_{k}\left[n\right]\geq0,\forall k\in\mathcal{K},\label{P1-C2}\\
& \quad \Delta\boldsymbol{\mu}_{k}\left[n\right]\in\boldsymbol{\mathcal{U}}_{k},\forall k\in\mathcal{K},\label{P1-C3}\\
& \quad \parallel\mathbf{q}\left[n\right]-\mathbf{q}\left[n-1\right]\parallel \leq V_{\textrm{max}}\Delta\tau,\label{P1-C4}\\
& \quad\mathbf{q}\left[n\right]\in\mathcal{D},\label{P1-C5}\\
& \quad \eqref{expected_control_equation},\label{P1-C6}
\end{align}
\end{subequations}
where constraints \eqref{P1-C1} and \eqref{P1-C2} indicate the transmit power budget, with $P_{\max}$ being the maximum transmit power of the network. Constraints \eqref{P1-C4} and \eqref{P1-C5} represent the drone’s maximum flight speed and flight boundary, with $V_{\textrm{max}}$ and $\mathcal{D}$ denoting the maximum allowable speed and the feasible flight region, respectively. Solving the optimization problem yields the optimal power allocation and drone trajectory at time step $n$, along with control inputs over the prediction horizon. However, it is worth mentioning that only the control input for the current slot is adopted. It is evident that problem \eqref{P1} is non-convex with coupled variables, making it difficult to achieve a globally optimal solution. To address this challenge, we develop an efficient iterative AO algorithm in the sequel.

\section{Proposed solutions \label{sec:Proposed-solutions}}

In this section, we first derive the closed-form expression of the outage probability function related to the transmit power and drone trajectory. Then, we reformulated the problem \eqref{P1} into a QP problem, which is next decomposed into three sub-problems and solved using an AO framework.
\begin{prop} \label{prop1}
The closed-form expression for $P_k[n]$ is given by 
\begin{equation}
 P_{k}\left[n\right] = 1 - \exp\left( - mg_k\left[n\right] \right) \sum_{i=0}^{m-1} \frac{1}{i!} \left( mg_k\left[n\right] \right)^i,\label{outage}            
\end{equation}
where $g_k\left[n\right]=d_{k}^{2}\left[n\right]\Gamma_{k}^{\textrm{th}}\sigma_{k}^{2} / \left( \alpha_0p_{k}\left[n\right]\right)$ with $\Gamma_{k}^{\mathrm{th}}>0$ denoting the minimum SNR required to support the target rate $R_{k}^{\mathrm{th}}$ at the $k$-th AGV.
\end{prop}
\begin{IEEEproof}
% See Appendix \ref{appendix1}.
The first derivative of $R_k\left(\Gamma_k[n]\right)$ is
    \begin{equation}
    \begin{split}
         R^{\prime}_k&\left(\Gamma_k[n]\right) =\frac{1}{1+\Gamma_k[n]}\\
         &\times \left(1- \frac{Q^{-1}\left(\epsilon\right)}{\sqrt{l_{k}}\left(1+\Gamma_k[n]\right)\sqrt{\left(1+\Gamma_k[n]\right)^2-1}}\right).
    \end{split}
    \end{equation}
    Setting $R^{\prime}_k\left(\Gamma_k[n]\right)=0$ yields
    \begin{equation}  \left(1+\Gamma_k[n]\right)\sqrt{\left(1+\Gamma_k[n]\right)^2-1} = \frac{Q^{-1}\left(\epsilon\right)}{\sqrt{l_{k}}}.
    \end{equation}
   The corresponding solution is
    \begin{equation}
        \Gamma_{k,0}[n] = \sqrt{\frac{1}{2}+\sqrt{\frac{1}{4}+\frac{\left(Q^{-1}\left(\epsilon\right)\right)^2}{l_{k}}}} -1.
    \end{equation}
    Note that $\left(1+\Gamma_k[n]\right)\sqrt{\left(1+\Gamma_k[n]\right)^2-1}$ is monotonically increasing with respect to $\Gamma_k[n]$. Therefore, $R_k\left(\Gamma_k[n]\right)$ is monotonically decreasing on $\left(0,\Gamma_{k,0}[n]\right) $, and monotonically increasing on $\left(\Gamma_{k,0}[n], +\infty\right) $. Moreover, since $R_k\left(0\right) = 0 $, it follows that $R_k\left(\Gamma_{k,0}[n]\right) < 0 $. Hence, $R_k\left(\Gamma_k[n]\right)$ is a monotonically increasing function of $\Gamma_k[n]$ over the feasible domain. According to \cite{analytical_solution}, the closed-form solution to $R_k\left(\Gamma_k[n]\right)=R_{k}^{\textrm{th}}$ is $\Gamma_k[n] = 2^{R_{k}^{\textrm{th}}+\frac{\kappa^*}{2}-1}$, where 
    \begin{equation}
\kappa^*=\mathcal{W}\left(^{2\frac{Q^{-1}\left(\epsilon\right)}{\sqrt{l_{k}}},-2\frac{Q^{-1}\left(\epsilon\right)}{\sqrt{l_{k}}}};-4\times2^{-2R_{k}^{\textrm{th}}}\frac{\left(Q^{-1}\left(\epsilon\right)\right)^2}{l_{k}}\right),  
    \end{equation}
    with $\mathcal{W}\left(^{\omega_1,\omega_2};\mu\right)$ defined as
    \begin{equation}
    \begin{split}
\mathcal{W}\left(^{\omega_1,\omega_2};\mu\right)  &= \omega_1 - \sum^{+\infty}_{i=1}  \frac{1}{i\times i !}\\ & \times \left(\frac{2^{-\omega_1}\mu i }{\omega_2-\omega_1}\right)\mathcal{B}_{i-1}\left(\frac{-2}{i\left(\omega_2-\omega_1\right)}\right) ,  
    \end{split}
    \end{equation}
    and
    \begin{equation}
    \mathcal{B}_{i}\left(z\right) = \sum^{i}_{j=0} \frac{\left(i+j\right)!}{j!\left(i-j\right)!}\left(\frac{z}{2}\right)^{j}.
    \end{equation}
    Thus, \eqref{outage1} can be equivalently expressed in terms of the SNR as
    \begin{equation}
P_{k}\left[n\right]=\Pr\left(\Gamma_{k}\left[n\right]<2^{R_{k}^{\textrm{th}}+\frac{\kappa^*}{2}-1}\right).\label{outage2}
    \end{equation}
    For convenience, define $\Gamma_{k}^{\mathrm{th}}=2^{R_{k}^{\textrm{th}}+\frac{\kappa^*}{2}-1}$ as the corresponding minimum SNR. Then, \eqref{outage2} can be rewritten as
    \begin{equation}
P_{k}\left[n\right]=\Pr\left(\Gamma_{k}\left[n\right]<\Gamma_{k}^{\textrm{th}}\right).\label{outage3}
\end{equation}
Substituting \eqref{SINR} into \eqref{outage3} yields the outage probability expression 
% \cite{los_nlos_independent}
\begin{equation}
    P_{k}\left[n\right]=\Pr\left(\lvert \hat{h}_{k}[n]\lvert^2 < \frac{\Gamma_{k}^{\textrm{th}}d^2_k[n]\sigma_{k}^{2}}{\alpha_0p_{k}\left[n\right]}\right).\label{outage4}
\end{equation}
Since $\hat{h}_{k}[n]$ follows the Nakagami-$m$ distribution, the real-time outage probability reduces to \eqref{outage}, which completes the proof \cite{nakagami5}.\footnote{\textcolor{black}{To facilitate analytical tractability and enable closed-form derivations, the fading parameter $m$ is set to a positive integer. Although this introduces a simplification of the geometry-dependent fading behavior, the impact on practical performance is negligible \cite{nakagami6}.}}  
\end{IEEEproof}
To facilitate tractable optimization, we proceed by transforming problem \eqref{P1} into an equivalent QP formulation. Given the system state $\tilde{\mathbf{x}}_{k}[n]$, the predicted control states over the horizon are expressed as \eqref{evolution_cs}, as shown at the top of the next page.
\begin{figure*}[tp]
\begin{equation}
\begin{split}\tilde{\mathbf{x}}_{k}\left[n\lvert n \right] & =\tilde{\mathbf{x}}_{k}\left[n\right],\\
\mathbb{E}\left[\tilde{\mathbf{x}}_{k}\left[n+1\lvert n \right]\right] & =\mathbf{\tilde{A}}_{k}\tilde{\mathbf{x}}_{k}\left[n\right]+\left(1-P_{k}\left[n\right]\right)\mathbf{\mathbf{\tilde{B}}}_{k} \Delta\mathbf{u}_{k}\left[n\lvert n \right],\\
\mathbb{E}\left[\tilde{\mathbf{x}}_{k}\left[n+2\lvert n\right]\right]&=\mathbf{\tilde{A}}_{k}\mathbb{E}\left[\tilde{\mathbf{x}}_{k}\left[n+1\lvert n\right]\right]+\left(1-P_{k}\left[n\right]\right)\mathbf{\mathbf{\tilde{B}}}_{k}\Delta\mathbf{u}_{k}\left[n+1\lvert n\right],\\&=\mathbf{\mathbf{\tilde{A}}}_{k}^{2}\tilde{\mathbf{x}}_{k}\left[n\right]+\mathbf{\mathbf{\tilde{A}}}_{k}\left(1-P_{k}\left[n\right]\right)\mathbf{\mathbf{\tilde{B}}}_{k}\Delta\mathbf{u}_{k}\left[n\lvert n\right]+\mathbf{\tilde{B}}_{k}\left(1-P_{k}\left[n\right]\right)\Delta\mathbf{u}_{k}\left[n+1\lvert n\right],\\
 & \vdots\\
\mathbb{E}\left[\tilde{\mathbf{x}}_{k}\left[n+N_{p}\lvert n \right]\right] & =\mathbf{\tilde{A}}_{k}^{N_{p}}\mathbf{\tilde{\mathbf{x}}}_{k}\left[n\right]+\mathbf{\tilde{A}}_{k}^{N_{p}-1}\left(1-P_{k}\left[n\right]\right)\mathbf{\tilde{B}}_{k} \Delta\mathbf{u}_{k}\left[n\lvert n \right]+\ldots+\left(1-P_{k}\left[n\right]\right)\mathbf{\mathbf{\tilde{B}}}_{k} \Delta\mathbf{u}_{k}\left[n+N_{p}-1\lvert n \right].
\end{split}
\label{evolution_cs}
\end{equation}
\rule[0.5ex]{1\textwidth}{0.4pt}
\end{figure*}
Next, we introduce the predicted control state vector
\begin{equation}
    \boldsymbol{\chi}_{k}\left[n\right]=\left[\begin{array}{c}
\tilde{\mathbf{x}}_{k}\left[n\lvert n\right]\\
\tilde{\mathbf{x}}_{k}\left[n+1\lvert n\right]\\
\vdots\\
\tilde{\mathbf{x}}_{k}\left[n+N_{p}\lvert n\right]
\end{array}\right]\in\mathbb{R}^{6N_{p}+6}.
\end{equation}
Accordingly, \eqref{evolution_cs} can be rewritten in a matrix form as
\begin{equation}
\mathbb{E}\left[\boldsymbol{\chi}_{k}\left[n\right]\right]=\mathbf{C}_{k}\tilde{\mathbf{x}}_{k}\left[n\right]+\left(1-P_{k}\left[n\right]\right)\mathbf{F}_{k} \Delta\boldsymbol{\mu}_{k}\left[n\right],\label{equation_matrix}
\end{equation}
where $\mathbf{C}_{k}\in\mathbb{R}^{\left(6N_{p}+6\right)\times6}$ and $\mathbf{F}_{k}\in\mathbb{R}^{\left(6N_{p}+6\right)\times2N_{p}}$ denote the augmented state transition and input matrices, respectively, with detailed expressions provided as
\begin{equation}
    \mathbf{C}_{k}=\left[\begin{array}{c}
\mathbf{I}_{6\times6}\\
\mathbf{\tilde{A}}_{k}\\
\vdots\\
\mathbf{\tilde{A}}_{k}^{N_{p}}
\end{array}\right],
\mathbf{F}_{k}=\left[\begin{array}{cccc}
\mathbf{0} &  \cdots & \mathbf{0}\\
\mathbf{\tilde{B}}_{k} &  \cdots & \mathbf{0}\\
\mathbf{\tilde{A}}_{k}\mathbf{\tilde{B}}_{k} &  \cdots & \mathbf{0}\\
\vdots &  \ddots & \vdots\\
\mathbf{\tilde{A}}_{k}^{N_{p}-1}\mathbf{\tilde{B}}_{k} &  \cdots & \mathbf{\tilde{B}}_{k}
\end{array}\right].
\end{equation}
Similarly, we define that
\begin{equation}
    \boldsymbol{\chi}_{k,\textrm{ref}}\left[n\right]=\left[\begin{array}{c}
\tilde{\mathbf{x}}_{k,\textrm{ref}}\left[n\right]\\
\tilde{\mathbf{x}}_{k,\textrm{ref}}\left[n+1\right]\\
\vdots\\
\tilde{\mathbf{x}}_{k,\textrm{ref}}\left[n+N_{p}\right]
\end{array}\right]\in\mathbb{R}^{6N_{p}+6},
\end{equation}
represents the reference state trajectory for the $k$-th AGV over the prediction horizon. The cost function in problem \eqref{P1} is equivalently rewritten in the matrix form as in \eqref{equation_matrix1}, shown at the top of the next page,
\begin{figure*}[tp]
\begin{equation}
\hat{J}_{n}=\sum_{k\in\mathcal{K}}\mathbb{E}\left[\left(\boldsymbol{\chi}_{k}\left[n\right]-\boldsymbol{\chi}_{k,\textrm{ref}}\left[n\right]\right)^{\textrm{T}}\mathbf{Q}_{4}\left(\boldsymbol{\chi}_{k}\left[n\right]-\boldsymbol{\chi}_{k,\textrm{ref}}\left[n\right]\right)+ \Delta\boldsymbol{\mu}_{k}\left[n\right]^{\textrm{T}}\mathbf{Q}_{5} \Delta\boldsymbol{\mu}_{k}\left[n\right]\right].\label{equation_matrix1}
\end{equation}
\rule[0.5ex]{1\textwidth}{0.4pt}
\end{figure*}
where $\mathbf{Q}_{4}=\textrm{diag}\left(\mathbf{Q}_{1},\ldots,\mathbf{Q}_{1},\mathbf{Q}_{3}\right)$ and $\mathbf{Q}_{5}=\textrm{diag}\left(\mathbf{Q}_{2},\ldots,\mathbf{Q}_{2}\right)$. By substituting \eqref{equation_matrix} into \eqref{equation_matrix1}, the cost function is rewritten as
\begin{equation}
\hat{J}_{n} = \sum_{k\in\mathcal{K}} \Delta{\boldsymbol{\mu}^{\textrm{T}}_{k}\left[n\right]}\mathbf{G}_{k}\left[n\right] \Delta\boldsymbol{\mu}_{k}\left[n\right]
 +\mathbf{H}_{k}\left[n\right] \Delta\boldsymbol{\mu}_{k}\left[n\right]+M_{k}\left[n\right],
\label{cost_matrix2}
\end{equation}
where $\mathbf{G}_{k}\left[n\right]$, $\mathbf{H}_{k}\left[n\right]$, $M_{k}\left[n\right]$ are constant coefficient, which are calculated as
% \begin{equation}
%     \begin{split}
%         \mathbf{G}_{k}\left[n\right]=&\left(1-P_{k}\left[n\right]\right)^{2}\mathbf{F}_{k}^{\textrm{T}}\mathbf{Q}_{4}\mathbf{F}_{k}+\mathbf{Q}_{5},\\
%         \mathbf{H}_{k}\left[n\right]=&2\left(1-P_{k}\left[n\right]\right)\left(\tilde{\mathbf{x}}_{k}\left[n\right]^{\textrm{T}}\mathbf{C}_{k}^{\textrm{T}}-\boldsymbol{\chi}_{k,\textrm{ref}}^{\textrm{T}}\left[n\right]\right)\mathbf{Q}_{4}\mathbf{F}_{k},\\
%         R_{k}\left[n\right]= & \tilde{\mathbf{x}}^{\textrm{T}}_{k}\left[n\right]\mathbf{C}_{k}^{\textrm{T}}\mathbf{Q}_{4}\mathbf{C}_{k}\mathbf{\tilde{\mathbf{x}}}_{k}\left[n\right]-2\mathbf{\tilde{\mathbf{x}}}^{\textrm{T}}_{k}\left[n\right]\mathbf{C}_{k}^{\textrm{T}}\mathbf{Q}_{4}\boldsymbol{\chi}_{k,\textrm{ref}}\left[n\right]\\
%  & +\boldsymbol{\chi}_{k,\textrm{ref}}^{\textrm{T}}\left[n\right]\mathbf{Q}_{4}\boldsymbol{\chi}_{k,\textrm{ref}}\left[n\right].
%     \end{split}
% \end{equation}
\begin{equation}
\mathbf{G}_{k}\left[n\right]=\left(1-P_{k}\left[n\right]\right)^{2}\mathbf{F}_{k}^{\textrm{T}}\mathbf{Q}_{4}\mathbf{F}_{k}+\mathbf{Q}_{5},
\end{equation}
\begin{equation}
\mathbf{H}_{k}\left[n\right]=2\left(1-P_{k}\left[n\right]\right)\left(\tilde{\mathbf{x}}_{k}\left[n\right]^{\textrm{T}}\mathbf{C}_{k}^{\textrm{T}}-\boldsymbol{\chi}_{k,\textrm{ref}}^{\textrm{T}}\left[n\right]\right)\mathbf{Q}_{4}\mathbf{F}_{k},
\end{equation}
\begin{equation}
\begin{split}M_{k}\left[n\right]= & \tilde{\mathbf{x}}^{\textrm{T}}_{k}\left[n\right]\mathbf{C}_{k}^{\textrm{T}}\mathbf{Q}_{4}\mathbf{C}_{k}\mathbf{\tilde{\mathbf{x}}}_{k}\left[n\right]-2\mathbf{\tilde{\mathbf{x}}}^{\textrm{T}}_{k}\left[n\right]\mathbf{C}_{k}^{\textrm{T}}\mathbf{Q}_{4}\boldsymbol{\chi}_{k,\textrm{ref}}\left[n\right]\\
 & +\boldsymbol{\chi}_{k,\textrm{ref}}^{\textrm{T}}\left[n\right]\mathbf{Q}_{4}\boldsymbol{\chi}_{k,\textrm{ref}}\left[n\right].
\end{split}
\end{equation}
Consequently, problem \eqref{P1} is reformulated as
\begin{subequations}\label{P2}
\begin{align}
\min_{\left\{\Delta\boldsymbol{\mu}_{k}\left[n\right]\right\},\mathbf{p}\left[n\right],\mathbf{q}\left[n\right]} & \quad \hat{J}_{n}\left(\left\{\Delta\boldsymbol{\mu}_{k}\left[n\right]\right\},\mathbf{p}\left[n\right],\mathbf{q}\left[n\right]\right)\\
\mathrm{s.t.} & \quad \eqref{P1-C1},\eqref{P1-C2},\eqref{P1-C3},\eqref{P1-C4},\eqref{P1-C5}.\label{P2-C1}
\end{align}
\end{subequations}
Although problem \eqref{P2} is cast as a QP problem, it remains challenging to solve due to the interdependence among control inputs, power allocation, and drone trajectories. To overcome this challenge, we decompose problem \eqref{P2} into three sub-problems: AGV path tracking, power allocation, and drone trajectory planning. These sub-problems are then solved in an alternating manner using the AO method.

\subsection{Control Strategy Design}
In this subsection, we focus on the control strategy design for AGV path tracking, assuming that the power allocation and drone trajectory are predetermined. The corresponding optimization problem is formulated as 
\begin{subequations}\label{P3}
\begin{align}
\min_{\left\{\Delta\boldsymbol{\mu}_{k}\left[n\right]\right\}} & \quad \hat{J}_{n}\left(\left\{\Delta\boldsymbol{\mu}_{k}\left[n\right]\right\}\right)\\
\mathrm{s.t.} 
& \quad \eqref{P1-C3}.\label{P3-C1}
\end{align}
\end{subequations}
Given the objective function $\hat{J}_{n}\left(\left\{\Delta\boldsymbol{\mu}_{k}\left[n\right]\right\}\right)$ is quadratic and involves a positive definite weighting matrix  $\mathbf{G}_{k}\left[n\right]$, problem \eqref{P3} can be solved as a standard QP problem using convex optimization techniques \cite{convex_QP1}, \cite{convex_optimization}. Furthermore, since only the control policy of the current interval is selected and transmitted to AGVs, the optimal control input $\Delta\mathbf{u}_{k}^{\star}\left[n\right]$ is obtained as the first two entries of the optimized control sequence $ \Delta\boldsymbol{\mu}_{k}^{\star}\left[n\right]$.

\subsection{Optimization of Power Allocation}

With the given drone trajectory and the updated control action, problem \eqref{P2} is equivalently written as
\begin{subequations}\label{P4}
\begin{align}
\min_{\mathbf{p}\left[n\right]} & \quad \hat{J}_{n}\left(\mathbf{p}\left[n\right]\right)\\
\mathrm{s.t.} & \quad \eqref{P1-C1},\eqref{P1-C2},\label{P4-C1}
\end{align}
\end{subequations}
where $\hat{J}_{n}\left(\mathbf{p}\left[n\right]\right)$ denotes the control cost associated with the power allocation variable and is given by
\begin{equation}
\begin{split}\hat{J}_{n}\left(\mathbf{p}\left[n\right]\right)= & \sum_{k\in\mathcal{K}}L_{k}\left[n\right]\left(\psi\left(p_k\left[n\right]\right)\right)^{2}\\
 & + 2S_{k}\left[n\right]\psi\left(p_k\left[n\right]\right)+W_{k}\left[n\right],
\end{split}
\end{equation}
with
\begin{align}
\psi\left(p_k\left[n\right]\right)=\exp\left( - mg\left(p_k\left[n\right]\right) \right) \sum_{i=0}^{m-1} \frac{1}{i!} \left( mg\left(p_k\left[n\right]\right) \right)^i,
\end{align}
\begin{align}
L_{k}\left[n\right]= \Delta\boldsymbol{\mu}_{k}^{\textrm{T}}\left[n\right]\mathbf{F}_{k}^{\textrm{T}}\mathbf{Q}_{4}\mathbf{F}_{k} \Delta\boldsymbol{\mu}_{k}\left[n\right],
\end{align}
\begin{align}
S_{k}\left[n\right]=\left(\tilde{\mathbf{x}}_{k}\left[n\right]^{\textrm{T}}\mathbf{C}_{k}^{\textrm{T}}-\boldsymbol{\chi}_{k,\textrm{ref}}^{\textrm{T}}\left[n\right]\right)\mathbf{Q}_{4}\mathbf{F}_{k} \Delta\boldsymbol{\mu}_{k}\left[n\right],
\end{align}
\begin{align}
W_{k}\left[n\right]= \Delta{\boldsymbol{\mu}}^{\textrm{T}}_{k}\left[n\right]\mathbf{Q}_{5} \Delta\boldsymbol{\mu}_{k}\left[n\right]+{M}_{k}[n].
\end{align}
Since problem \eqref{P4} is a linearly constrained optimization problem with a non-convex objective function, it can be solved through the PGD method \cite{PGD}. The detailed procedure of the PGD-based algorithm for power allocation is summarized in Algorithm \ref{alg:pgd}. In Algorithm \ref{alg:pgd}, the process starts by initializing a feasible starting point, assuming equal power allocation among all AGVs. The initial step size $\rho$, Armijo parameter $\beta$, backtracking coefficient $\hat{\rho}$, and convergence tolerance $\varepsilon$ are also specified. At each iteration, the search direction is determined by computing the gradient of $\hat{J}_{n}\left(\mathbf{p}\left[n\right]\right)$ with respect to $\mathbf{p}\left[n\right]$, i.e.,
\begin{equation}
   \boldsymbol{\nabla} _{\mathbf{p}\left[n\right]} \hat{J}_{n}\left(\mathbf{p}\left[n\right]\right) = \left[\frac{\partial \hat{J}_{n}\left(\mathbf{p}\left[n\right]\right)}{\partial p_{1}\left[n\right]}
   ,\ldots,\frac{\partial \hat{J}_{n}\left(\mathbf{p}\left[n\right]\right)}{\partial p_{K}\left[n\right]}
   \right]^{\textrm{T}},\label{gradient}
\end{equation}
In \eqref{gradient}, $\frac{\partial \hat{J}_{n}\left(\mathbf{p}\left[n\right]\right)}{\partial p_{k}\left[n\right]}$ is expressed as
\begin{equation}
\begin{split}
 \frac{\partial \hat{J}_{n}\left(\mathbf{p}\left[n\right]\right)}{\partial p_{k}\left[n\right]} = & 2L_{k}\left[n\right]\psi\left(p_k\left[n\right]\right) \\
\times&\frac{\partial \psi\left(p_k\left[n\right]\right)}{\partial p_{k}\left[n\right]}+2S_{k}\left[n\right]\frac{\partial \psi\left(p_k\left[n\right]\right)}{\partial p_{k}\left[n\right]},
\end{split}
\end{equation}
where $\frac{\partial \psi\left(p_k\left[n\right]\right)}{\partial p_{k}\left[n\right]}$ is particularized as 
\begin{equation}
\begin{split}
    \frac{\partial \psi\left(p_k\left[n\right]\right)}{\partial p_{k}\left[n\right]}&=\frac{md_{k}^{2}\left[n\right]\Gamma_{k}^{\textrm{th}}\sigma_{k}^{2}}{\left(m-1\right)!\alpha_0p^2_{k}\left[n\right]}\\
    &\times    \exp\left(-mg\left(p_k\left[n\right]\right)\right)\left(mg\left(p_k\left[n\right]\right)\right)^{m-1}.
\end{split}
\end{equation}
Then, $\mathbf{p}\left[n\right]$ is updated by projecting onto the feasible domain, i.e, $\Psi_{\mathcal{P}}\left(\mathbf{p}^{\prime}\left[n\right]-\rho\boldsymbol{\Delta}\right):\mathbb{R}^{K}\rightarrow\mathcal{P}$, where $\mathcal{P}$ denotes the feasible set defined as
\begin{equation}
    \mathcal{P}=\left\{ \mathbf{p}\left[n\right]\lvert\sum_{k\in\mathcal{K}}p_{k}\left[n\right]\leq P_{\max},p_{k}\left[n\right]\geq0\right\},
\end{equation}
and $\Psi_{\mathcal{P}}\left(\mathbf{p}^{\prime}\left[n\right]-\rho\boldsymbol{\Delta}\right)$ is obtained via solving problem
 \begin{subequations} \label{P44}
    \begin{align}
\min_{\mathbf{p}\left[n\right]\in\mathcal{P}} & \quad\parallel\mathbf{p}^{\prime}\left[n\right]-\rho\boldsymbol{\Delta}-\mathbf{p}\left[n\right]\parallel.\label{P44-1}
\end{align}
\end{subequations}
The step size $\rho$ is further adaptively selected via a backtracking line search to ensure sufficient descent \cite{PGD_bsl}. Specifically, starting from an initial value, $\rho$ is iteratively reduced by a factor of $1/(1+\hat{\rho})$ until the following Armijo-type condition is satisfied
\begin{equation}
    \hat{J}_{n}\left(\Psi_{\mathcal{P}}\left(\mathbf{p}'[n] - \rho \boldsymbol{\Delta}\right)\right)
    \le \hat{J}_{n}\left(\mathbf{p}'[n]\right) - \beta \rho \left\|\boldsymbol{\Delta}\right\|^2.
    \label{eq:armijo}
\end{equation}
\begin{algorithm}[t]
\caption{PGD-Based Algorithm for Power Allocation\label{alg:pgd}}

1: \textbf{Initialize} Starting feasible point $\mathbf{p}\left[n\right]\leftarrow\left[\frac{P_{\max}}{K},\ldots,\frac{P_{\max}}{K}\right]$, 

$\quad\qquad\qquad$ initial step size $\rho$, Armijo parameter $\beta$, backtrack-

$\quad\qquad\qquad$ ing coefficient $\hat{\rho}$, and tolerance $\varepsilon$.

2: \textbf{repeat}

3: $\qquad$Save the previous direction vector $\mathbf{p}^{\prime}\left[n\right]\leftarrow\mathbf{p}\left[n\right]$.

4: $\qquad$Determine a search direction $\boldsymbol{\Delta}\triangleq\boldsymbol{\nabla} _{\mathbf{p}\left[n\right]} \hat{J}_{n}\left(\mathbf{p}\left[n\right]\right)$.

5: $\qquad$\textbf{repeat}

6: $\qquad$$\qquad$Choose a step size $\rho\leftarrow\frac{\rho}{1+
\hat{\rho}}$.

7: $\qquad$\textbf{until} $\hat{J}_{n}\left(\Psi_{\mathcal{P}}\left(\mathbf{p}'[n] - \rho \boldsymbol{\Delta}\right)\right)
    \le \hat{J}_{n}\left(\mathbf{p}'[n]\right) - \beta \rho \left\|\boldsymbol{\Delta}\right\|^2$.

8: $\qquad$Update $\mathbf{p}\left[n\right]\leftarrow\Psi_{\mathcal{P}}\left(\mathbf{p}^{\prime}\left[n\right]-\rho\boldsymbol{\Delta}\right)$.

9: \textbf{until} $\parallel\mathbf{p}\left[n\right]-\mathbf{p}^{\prime}\left[n\right]\parallel<\varepsilon$

10: \textbf{Output} $\mathbf{p}^{\star}\left[n\right]\leftarrow\mathbf{p}\left[n\right]$.
\end{algorithm}

\subsection{Drone Trajectory Design and Overall Algorithm}
In this subsection, we focus on optimizing the drone trajectory given the control strategy and power allocation obtained from previous steps. The corresponding optimization is reduced to
\begin{subequations}\label{P5}
\begin{align}
\min_{\mathbf{q}\left[n\right]} & \quad \hat{J}_{n}\left(\mathbf{q}\left[n\right]\right)\\
\mathrm{s.t.} & \quad \eqref{P1-C4},\eqref{P1-C5}.\label{P5-C1}
\end{align}
\end{subequations}
Due to the non-convexity of the objective function in problem \eqref{P5}, the SCA method is adopted, which linearizes the non-convex terms using first-order Taylor expansions \cite{SCA_CC}. In particular, $\hat{J}_{n}\left(\mathbf{q}\left[n\right]\right)$ is approximated as
\begin{equation}
    \phi= \hat{J}_{n}\left(\mathbf{q}^{\iota_1}\left[n\right]\right)+ \left[\boldsymbol{\nabla}_{\mathbf{q}\left[n\right]}\hat{J}_{n}\left(\mathbf{q}^{\iota_1}\left[n\right]\right)\right]^{\textrm{T}}\left(\mathbf{q}\left[n\right]-\mathbf{q}^{\iota_1}\left[n\right]\right),\label{SCA}
\end{equation}
where $\mathbf{q}^{\iota_1}\left[n\right]$ denotes the drone’s trajectory at the $n$-th time slot in the $\iota_1$-th iteration and $\boldsymbol{\nabla}_{\mathbf{q}\left[n\right]}\hat{J}_{n}\left(\mathbf{q}^{\iota_1}\left[n\right]\right)$ is given by \eqref{grad_q}, as shown at the top of the next page.
% , with $\partial \psi_{k}\left(\mathbf{q}^{\iota_1}\left[n\right]\right) /{\partial \mathbf{q}^{\iota_1}\left[n\right]}$ representing the first-order derivation of   $_{k}\left(\mathbf{q}^{\iota_1}\left[n\right]\right)$.
\begin{figure*}[tp]
\begin{equation}
\boldsymbol{\nabla}_{\mathbf{q}^{\iota_1}\left[n\right]}\hat{J}_{n}\left(\mathbf{q}^{\iota_1}\left[n\right]\right) =\sum_{k\in\mathcal{K}}2L_{k}\left[n\right]\psi_{k}\left(\mathbf{q}^{\iota_1}\left[n\right]\right) \frac{\partial \psi_{k}\left(\mathbf{q}^{\iota_1}\left[n\right]\right)}{\partial 
\mathbf{q}^{\iota_1}\left[n\right]}+2S_{k}\left[n\right]\frac{\partial \psi_{k}\left(\mathbf{q}^{\iota_1}\left[n\right]\right)}{\partial \mathbf{q}^{\iota_1}\left[n\right]}.\label{grad_q}
\end{equation}
\hrule
\end{figure*}
As a result, problem \eqref{P5} is approximated by a sequence of convex optimization problems
\begin{subequations}\label{P6}
\begin{align}
\min_{\mathbf{q}\left[n\right]} & \quad \phi\\
\mathrm{s.t.} & \quad \eqref{P1-C4},\eqref{P1-C5},\label{P6-C1}
\end{align}
\end{subequations}
which can be solved with standard convex optimization tools such as CVX \cite{CVX}. 

Consequently, we propose an AO-based algorithm that alternately optimizes the control strategy, power allocation, and drone trajectory. The pseudocode of the AO-based algorithm is given in Algorithm \ref{alg:AO}. Specifically, the algorithm begins by initializing \( \mathbf{p}[n] \), \( \mathbf{q}[n] \), and $\tilde{\mathbf{x}}_{k}\left[n\right]$, as well as the auxiliary parameters including the iteration index \( \iota \), the convergence tolerance \( \varepsilon \), and the maximum number of inner iterations \( \iota_{\max} \). The optimization variables of problem \eqref{P2} are then partitioned into three blocks, i.e., $\left\{\Delta\boldsymbol{\mu}_{k}\left[n\right]\right\}$, $\mathbf{p}\left[n\right]$, and $\mathbf{q}\left[n\right]$. We optimize the three partitioned variables alternately by solving problems \eqref{P3}, \eqref{P4}, and \eqref{P5}, respectively, while keeping the other variables fixed. It is important to note that the solutions to problems \eqref{P4} and \eqref{P5} are only near-optimal approximations to the original problem. Therefore, it is essential to analyze the convergence of the AO-based problem. For convergence, let $\eta\left(\left\{\Delta\boldsymbol{\mu}_{k}\left[n\right]\right\},\mathbf{p}\left[n\right],\mathbf{q}\left[n\right]\right)$, $\eta_1\left(\left\{\Delta\boldsymbol{\mu}_{k}\left[n\right]\right\},\mathbf{p}\left[n\right],\mathbf{q}\left[n\right]\right)$, $\eta_2\left(\left\{\Delta\boldsymbol{\mu}_{k}\left[n\right]\right\},\mathbf{p}\left[n\right],\mathbf{q}\left[n\right]\right)$, 
and $\eta_3\left(\left\{\Delta\boldsymbol{\mu}_{k}\left[n\right]\right\},\mathbf{p}\left[n\right],\mathbf{q}\left[n\right]\right)$ denote the value of the objective function of problem \eqref{P2}, \eqref{P3}, \eqref{P4}, and \eqref{P6}, respectively. 
Accordingly, we have the following relationships \eqref{convergence}, as shown at the top of the next page, where the inequalities $1$, $2$, and $3$ hold as the solutions obtained by solving these three problems sequentially. 
\begin{figure*}[tp]
\begin{equation}
\begin{split}
    \eta\left(\left\{\Delta\boldsymbol{\mu}^{\iota-1}_{k}\left[n\right]\right\},\mathbf{p}^{\iota-1}\left[n\right],\mathbf{q}^{\iota-1}\left[n\right]\right) &  = \eta_1\left(\left\{\Delta\boldsymbol{\mu}^{\iota-1}_{k}\left[n\right]\right\},\mathbf{p}^{\iota-1}\left[n\right],\mathbf{q}^{\iota-1}\left[n\right]\right) \overset{1}{\geq} \eta_1\left(\left\{\Delta\boldsymbol{\mu}^{\iota}_{k}\left[n\right]\right\},\mathbf{p}^{\iota-1}\left[n\right],\mathbf{q}^{\iota-1}\left[n\right]\right)\\
    & \overset{2}{\geq} \eta_2\left(\left\{\Delta\boldsymbol{\mu}^{\iota}_{k}\left[n\right]\right\},\mathbf{p}^{\iota}\left[n\right],\mathbf{q}^{\iota-1}\left[n\right]\right) \overset{3}{\geq} \eta_3\left(\left\{\Delta\boldsymbol{\mu}^{\iota}_{k}\left[n\right]\right\},\mathbf{p}^{\iota}\left[n\right],\mathbf{q}^{\iota}\left[n\right]\right)\\
    & =\eta\left(\left\{\Delta\boldsymbol{\mu}^{\iota}_{k}\left[n\right]\right\},\mathbf{p}^{\iota}\left[n\right],\mathbf{q}^{\iota}\left[n\right]\right). \label{convergence}
    \end{split}
\end{equation}
\hrule
\end{figure*}
Equation \eqref{convergence} shows that the objective function of problem \eqref{P2} is non-increasing with each iteration. Moreover, as the objective function is lower bounded, Algorithm \ref{alg:AO} is guaranteed to converge to a sub-optimal solution.

We further analyze the computational complexity of the proposed AO-based algorithm. Problem \eqref{P3} is addressed using a standard interior-point method, with a computational complexity of $C_{1}=\mathcal{O}\left(\left(2N_{p}K\right)^{3.5}\right)$ \cite{SCA_CC}. The PGD-based algorithm incurs complexity mainly from gradient computations of  $\hat{J}_{n}\left(\mathbf{p}\left[n\right]\right)$ with respect to $\mathbf{p}\left[n\right]$ at each iteration, yielding a total complexity of $C_{2}=\mathcal{O}\left(\log\left(1/\varepsilon^{2}\right)K^{2}\right)$ \cite{PGD_CC}. The drone trajectory is optimized via the interior-point method combined with the SCA approach, which introduces a complexity of $C_{3}=\mathcal{O}\left(L_{3}\left(3M\right)^{3.5}\right)$, where $L_3$ denotes the number of SCA iterations. Therefore, the overall computational complexity of Algorithm \ref{alg:AO} is given $C_{tot}=\mathcal{O}\left(L\left(C_{1}+C_{2}+C_{3}\right)\right)$, with $L$ denoting the total number of alternating iterations.

\begin{algorithm}
\caption{AO-based Algorithm for Solving Original Problem \eqref{P2} \label{alg:AO}}

1: \textbf{Initialize} $n\leftarrow1$, $\iota\leftarrow1$, $\mathbf{p}^{\iota}\left[n\right]$, $\mathbf{q}^{\iota}\left[n\right]$,
$\tilde{\mathbf{x}}_{k}\left[n\right]$, tolerance

$\qquad\quad\qquad$ $\varepsilon$, and maximum number of iterations $\iota_{\max}$.

2: \textbf{repeat}

3: \textbf{$\qquad$repeat}

4: $\qquad$$\qquad$$\iota\leftarrow\iota+1$.

5: $\qquad$\textbf{$\qquad$}Solve problem \eqref{P3} with given $\tilde{\mathbf{x}}_{k}\left[n\right]$, $\mathbf{p}^{\iota-1}\left[n\right]$, 

$\qquad\quad\qquad$ and $\mathbf{q}^{\iota-1}\left[n\right]$ to obtain $\left\{\Delta\boldsymbol{\mu}^{\iota}_{k}\left[n\right]\right\}$.

6: $\qquad$\textbf{$\qquad$}Solve problem \eqref{P4} using Algoithm \ref{alg:pgd}, $\tilde{\mathbf{x}}_{k}\left[n\right]$, 

$\qquad\quad\qquad$ $\left\{\Delta\boldsymbol{\mu}^{\iota}_{k}\left[n\right]\right\}$, and $\mathbf{q}^{\iota-1}\left[n\right]$ to obtain $\mathbf{p}^{\iota}\left[n\right]$.

7: $\qquad$\textbf{$\qquad$}Solve problem \eqref{P6} using SCA, $\tilde{\mathbf{x}}_{k}\left[n\right]$, 

$\qquad\quad\qquad$ $\left\{\Delta\boldsymbol{\mu}^{\iota}_{k}\left[n\right]\right\}$, and $\mathbf{p}^{\iota}\left[n\right]$ to obtain $\mathbf{q}^{\iota}\left[n\right]$.

8: \textbf{$\qquad$until} The fractional increase of the objective value is 

$\qquad\quad\qquad\ $ below $\varepsilon$ or $\iota>\iota_{\max}$.

9: \textbf{$\qquad$Output }$ \Delta\mathbf{u}_{k}^{\star}\left[n\right]\leftarrow\left( \Delta\boldsymbol{\mu}_{k}^{\iota}\left[n\right]\right)_{1},\mathbf{p}^{\star}\left[n\right]\leftarrow\mathbf{p}^{\iota}\left[n\right]$, 

$\qquad\qquad\qquad\ $ and $\mathbf{q}^{\star}\left[n\right]\leftarrow\mathbf{q}^{\iota}\left[n\right]$.

10: \textbf{$\quad\ $} Use $\tilde{\mathbf{x}}_{k}\left[n\right]$ and $\Delta\mathbf{u}_{k}^{\star}\left[n\right]$ to obtain $\tilde{\mathbf{x}}_{k}\left[n+1\right]$.

11: $\qquad$$n\leftarrow n+1$.

12: \textbf{until} $n>N$.
\end{algorithm}

\section{ Results Discussion\label{sec:4}}

% \begin{figure}[t]
% \begin{table}[H]
% \caption{\label{tab:Table-of-parameters}Table of parameters}

% \centering{}%
% \begin{tabular}{|c|c|c|c|}
% \hline 
% $K$ & $3$ & $\sigma_{k}^{2}$ & $-100\ \textrm{dBm}$\tabularnewline
% \hline 
% $T$ & $60\ \textrm{s}$ & $R_k$ & $10\ \textrm{bps/Hz}$\tabularnewline
% \hline 
% $N$ & $120$ & $\mathbf{Q}_{k}$ & $\mathbf{I}_{6\times6}$\tabularnewline
% \hline 
% $\alpha_{0}$ & $-50\ \textrm{dB}$ & $\mathbf{R}_{k}$ & $\mathbf{I}_{2\times2}$\tabularnewline
% \hline 
% \end{tabular}
% \end{table}
% \end{figure}

\subsection{Simulation Results}

In this subsection, we present simulation results for the proposed algorithm, which jointly optimizes the control strategy, power allocation, and drone trajectory in a LAWN for real-time AGV path tracking. The total simulation duration is set to $60\ \textrm{s}$, divided into $120$ discrete time slots. Four AGVs are assumed to move within a predefined rectangular area of size $200\times200\ \textrm{m}^{2}$. The drone flies within the same area at a fixed altitude of $50\ \textrm{m}$. The initial drone position is set as  $\mathbf{q}\left[1\right]=\left[0,100\right]^{\textrm{T}} \ \textrm{m}$. Regarding the communication parameters, the Nakagami shape parameter is set to $m=2$. The reference channel gain is set to $\alpha_0 = -50\ \textrm{dB}$ and the sub-channel bandwidth $B$ is set to $1\ \textrm{MHz}$. The noise power is configured as $\sigma_k^2 = -100\ \textrm{dBm}$ and BLER is set to $\epsilon = 10^{-6}$ \cite{UAV_S&C3}, \cite{BLER}. For the control settings, the weighting matrices $\mathbf{Q}_1$ and $\mathbf{Q}_2$ are both set to identity matrices, and the prediction horizon is set to $N_p = 10$ \cite{MPC_increment}, \cite{WNCS_MPC}. 
\textcolor{black}{To assess the effectiveness of the proposed algorithm, four representative benchmark schemes are introduced for comparison. \emph{(a)} \textbf{Predictive PID}: Predictive PID extends the conventional PID framework by incorporating model-based prediction over a finite horizon to jointly optimize the control strategy, power allocation, and drone trajectory.
\emph{(b)} \textbf{Predictive LQR}: Predictive LQR controller with a fixed feedback gain integrates predictive optimization to dynamically adjust power allocation and drone trajectory. \textit{(c)} \textbf{Equal power allocation (EPA)}: The total power budget $P_{\max}$ is uniformly distributed among all AGVs, without dynamic power reallocation. \textit{(d)} \textbf{Drone-straight flight (SF)}: The drone follows a pre-defined straight-line trajectory with a randomly selected constant speed, without trajectory optimization.}
% To evaluate the performance of the proposed algorithm, we further introduce four benchmark schemes for comparison. The first is the equal power allocation (EPA) scheme, in which the total power budget $P_{\max}$ is equally distributed among all AGVs without dynamic power adjustment. The second is the drone-straight flight (SF) scheme, where the drone follows a pre-defined straight-line trajectory with a randomly selected constant speed, without trajectory optimization. 
Throughout the simulations, we analyze the convergence behavior, tracking performance, and drone trajectory under varying power budgets, rate thresholds, and command signal blocklengths.

\begin{figure}
\centering
\includegraphics[width=0.9\linewidth]{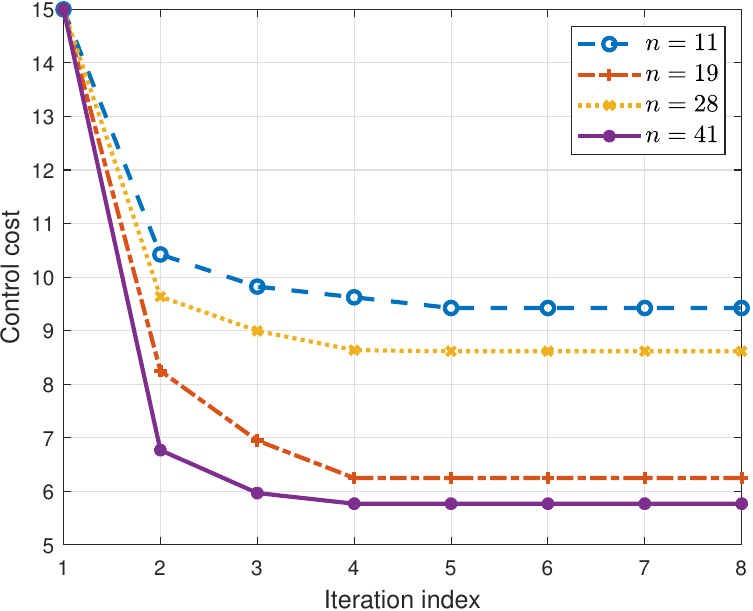}
\caption{\label{fig: Convergence}Convergence of the proposed algorithm at different time slots with $P_{\textrm{max}}=0\ \textrm{dBW}$, $R_k^{\textrm{th}}=1 \textrm{Mbps}$, and $l_k=1024\ \textrm{bits}$.}
\end{figure}

To evaluate the convergence behavior of the proposed algorithm, Fig. \ref{fig: Convergence} shows the evolution of control cost across iterations for different time slots. Specifically, four representative time slots, i.e., $n=11,19,28,41$, are selected to comprehensively assess the stability and convergence properties of the algorithm. As shown in the figure, the control cost in all cases rapidly decreases and converges to an optimal value within a few iterations, demonstrating both the rapid convergence rate and the effectiveness of the proposed optimization framework.

\begin{figure}
\centering
\includegraphics[width=0.9\linewidth]{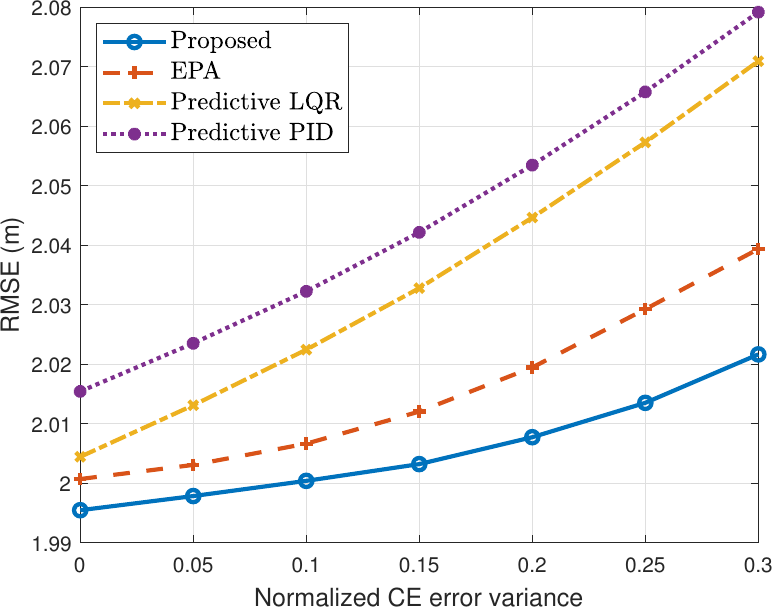}
\caption{\label{fig: CE}\textcolor{black}{Control performance versus normalized CE error variance under different benchmarks with $n=11$, $P_{\textrm{max}}=0\ \textrm{dBW}$, $R_k^{\textrm{th}}=1 \textrm{Mbps}$, and $l_k=1024\ \textrm{bits}$.}}
\end{figure}

\textcolor{black}{To further examine the robustness of the proposed method for the LAWN-enabled wireless control system, Fig. \ref{fig: CE} presents the tracking root mean square error (RMSE) versus the normalized channel estimation (CE) error variance under the proposed scenario and the benchmark methods. We adopt a minimum mean square error (MMSE) CE channel, where the CE error is modeled as a zero-mean complex Gaussian variable with variance $\sigma^2_{\mathrm{CE}}$ \cite{CE1}. The normalized CE error variance is defined as $\hat{\sigma}^2_{\mathrm{CE},k}=\sigma^2_{\mathrm{CE}}/\lvert h_{k}\left[n\right]\lvert^2,\forall k\in\mathcal{K}$. The results show that all schemes suffer increased RMSE as $\hat{\sigma}^2_{\mathrm{CE},k}$ grows, reflecting the degradation caused by imperfect channel state information (CSI). Nevertheless, the proposed algorithm consistently attains the lowest RMSE and the slowest growth rate, indicating that it can preserve accurate trajectory tracking even under pronounced CE uncertainty. Compared with the EPA scheme, which allocates power uniformly and cannot adapt to the reliability loss caused by poor channel estimation, the proposed design leverages joint optimization of power allocation, control actions, and drone trajectory, thereby mitigating the impact of CE errors. The predictive LQR and predictive PID controllers present higher RMSE and a steeper degradation, as they lack a unified optimization of control processes and are more sensitive to CE. Based on these observations, the subsequent simulations adopt the perfect CSI assumption and focus on the impact of transmission reliability and other system parameters.}

\begin{figure}
\centering
\includegraphics[width=0.9\linewidth]{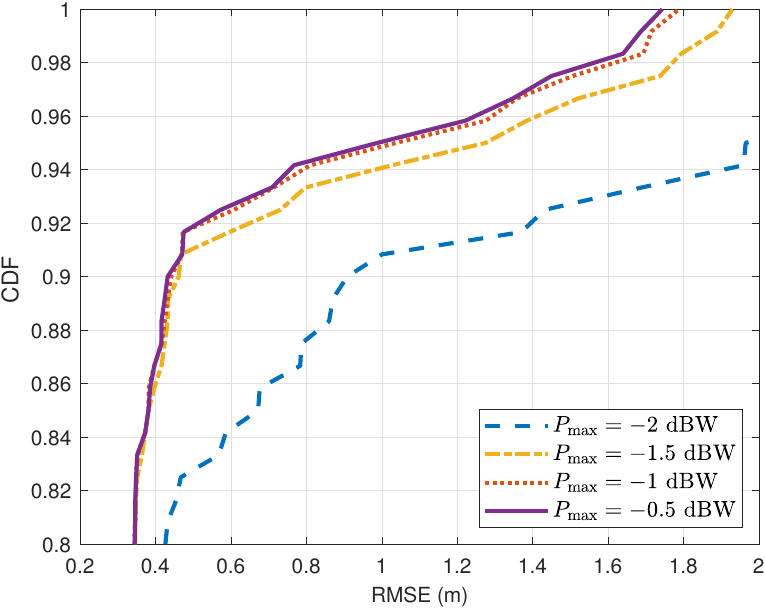}
\caption{\label{fig: CDF}CDF  of tracking performance under different power budgets with $R_k^{\textrm{th}}=1 \textrm{Mbps}$ and $l_k=1024\ \textrm{bits}$.}
\end{figure}

Fig.~\ref{fig: CDF} presents the cumulative distribution function (CDF) of the tracking root mean square error (RMSE) under various transmission power levels. The figure indicates that higher power budgets lead to improved control accuracy, as evidenced by smaller RMSE values and sharper CDF transitions. As the power budget decreases, performance degradation becomes more pronounced, especially in low-reliability scenarios, which underscores the strong interplay between wireless communication quality and control performance in real-time systems. The finding emphasizes that insufficient communication resources can significantly impair control accuracy, highlighting the need to jointly consider communication constraints when designing wireless control systems.

% \begin{figure}
% \centering
% \includegraphics[width=0.9\linewidth]{a_plot_reference/Baseline-eps-converted-to.pdf}
% \caption{\label{fig:Baseline}\textcolor{black}{Control performance versus power budget under different benchmarks with $n=11$, $R_k^{\textrm{th}}=1 \textrm{Mbps}$, and $l_k=1024\ \textrm{bits}$.}}
% \end{figure}

% \textcolor{black}{Fig. \ref{fig:Baseline} illustrates the control cost of the proposed algorithm compared with the baseline schemes under different transmit power budgets. As the power budget increases, all schemes exhibit improved control performance due to enhanced communication reliability and reduced packet loss. It is evident that the proposed algorithm consistently achieves the lowest control cost across all transmit power levels, demonstrating its superior efficiency in jointly optimizing power allocation, control actions, and drone trajectory. Different from the EPA scheme, which fails to adapt to variations in power availability, the proposed method dynamically redistributes power among AGVs according to instantaneous control demands, thereby achieving a better performance of tracking accuracy. In comparison, the predictive LQR and predictive PID controllers yield higher control costs, as they lack a unified optimization of control processes. Furthermore, the performance gap between the proposed method and the baselines widens at lower transmit power levels, revealing the strong robustness of our design in resource-constrained scenarios.}

\begin{figure}
\centering
\includegraphics[width=0.9\linewidth]{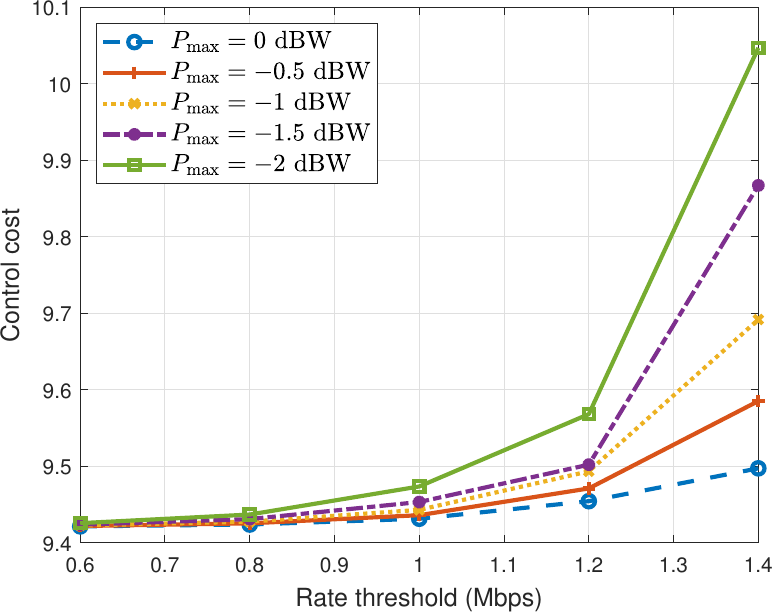}
\caption{\label{fig: control_R}Control cost versus the rate threshold under different power budgets with $n=11$ and $l_k=1024\ \textrm{bits}$.}
\end{figure}

\begin{figure}
\centering
\includegraphics[width=0.9\linewidth]{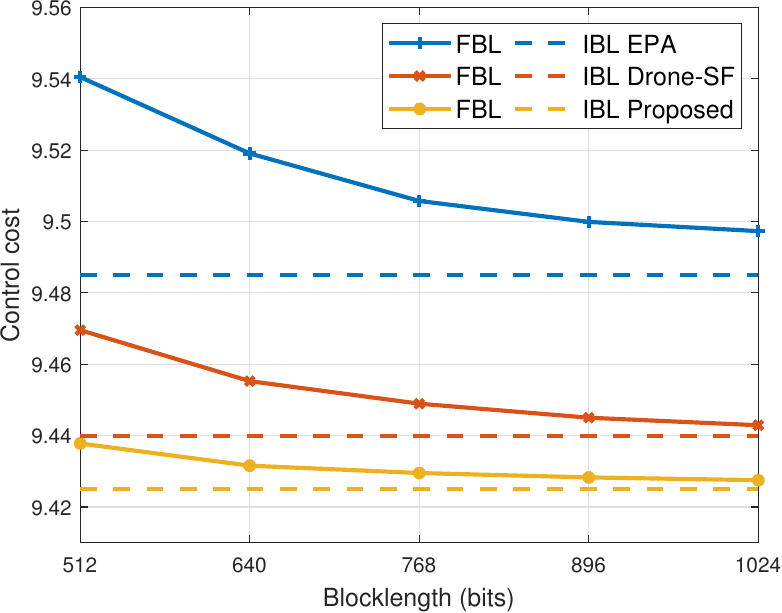}
\caption{\label{fig: blocklength}Control cost versus bloclength of command signals under different scenarios with $n=11$, $P_{\textrm{max}}=0\ \textrm{dBW}$, and $R^{\textrm{th}}_k=1\ \textrm{Mbps}$.}
\end{figure}

To further investigate the impact of wireless communication on control performance, Fig.~\ref{fig: control_R} shows the control cost at a selected time slot under different rate thresholds and power budgets. As the rate threshold increases, implying more stringent communication requirements, the control cost rises accordingly, which is because a higher rate threshold leads to a higher outage probability under limited power, reducing the reliability of command transmission and ultimately increasing the control cost. Moreover, the increase in control cost is more pronounced under lower power budgets since limited transmission power makes the system more sensitive to rate constraints, resulting in a sharper increase in outage probability and, consequently, higher control cost.

% \begin{figure}
%     \centering
%     % 第一张子图
%     \begin{subfigure}{0.9\linewidth}
%         \centering
%         \includegraphics[width=\linewidth]{a_plot_reference/path1-eps-converted-to.pdf}
%         \caption{\label{UPa}Proposed.} % 仅显示 (a)
%     \end{subfigure}
%     \\
%     \begin{subfigure}{0.9\linewidth}
%         \centering
%         \includegraphics[width=\linewidth]{a_plot_reference/path2-eps-converted-to.pdf}
%         \caption{\label{UPb} EPA.} % 仅显示 (a)
%     \end{subfigure}
%     % \hfill
%     % \begin{subfigure}{0.31\linewidth}
%     %     \centering
%     %     \includegraphics[width=\linewidth]{a_plot_reference/path3-eps-converted-to.pdf}
%     %     \caption{AGV 3} % 仅显示 (a)
%     % \end{subfigure}
%     \captionsetup{justification=raggedright} % 设置标题靠左对齐
%     \caption{\label{fig: Tracking path} Drone trajectory and AGVs tracking paths under different scenarios with $l_k=1024\ \textrm{bits}$, $P_{\textrm{max}}=0\ \textrm{dBW}$,  and $R^{\textrm{th}}_k=1\ \textrm{Mbps}$. }
% \end{figure}

Fig. \ref{fig: blocklength} illustrates the relationship between control cost and the blocklength of command signals under different benchmark scenarios. As expected, the control cost consistently decreases with increasing blocklength and gradually converges to the IBL case. This trend can be attributed to the improved communication reliability and reduced outage probability enabled by longer transmissions, which in turn enhance the quality of real-time wireless control. It is also evident that the proposed algorithm significantly outperforms the EPA scheme. This performance gain stems from the proposed algorithm's ability to dynamically adjust power allocation according to channel conditions and control demands, while the EPA baseline statically divides total power equally among AGVs, disregarding their heterogeneous needs and communication states. Likewise, the proposed method achieves better control performance than the drone SF scenario, where the drone follows a fixed flight path, limiting its ability to maintain reliable and timely communication.
\begin{figure*}
    \centering
    % 第一张子图
        \begin{subfigure}{0.32\linewidth}
        \centering
        \includegraphics[width=\linewidth]{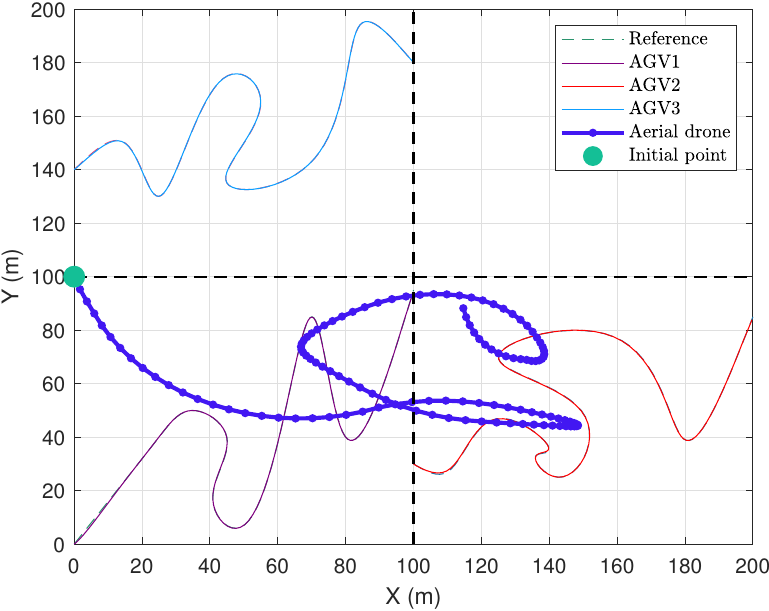}
        \caption{\label{UPc}Simulation with $3$ AGVs, Proposed.} % 仅显示 (a)
    \end{subfigure}
    \hfill
    \begin{subfigure}{0.32\linewidth}
        \centering
        \includegraphics[width=\linewidth]{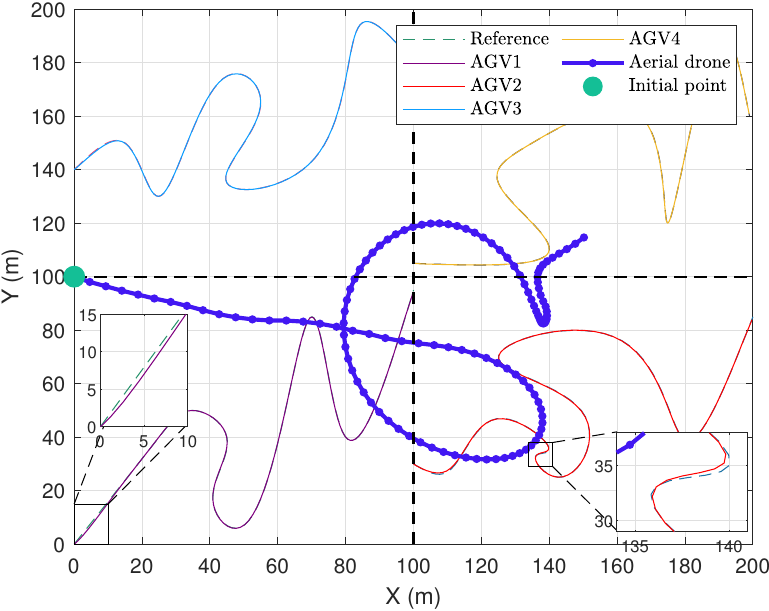}
        \caption{ \label{UPa}Simulation with $4$ AGVs, Proposed.} % 仅显示 (a)
    \end{subfigure}
    \hfill
    \begin{subfigure}{0.32\linewidth}
        \centering
        \includegraphics[width=\linewidth]{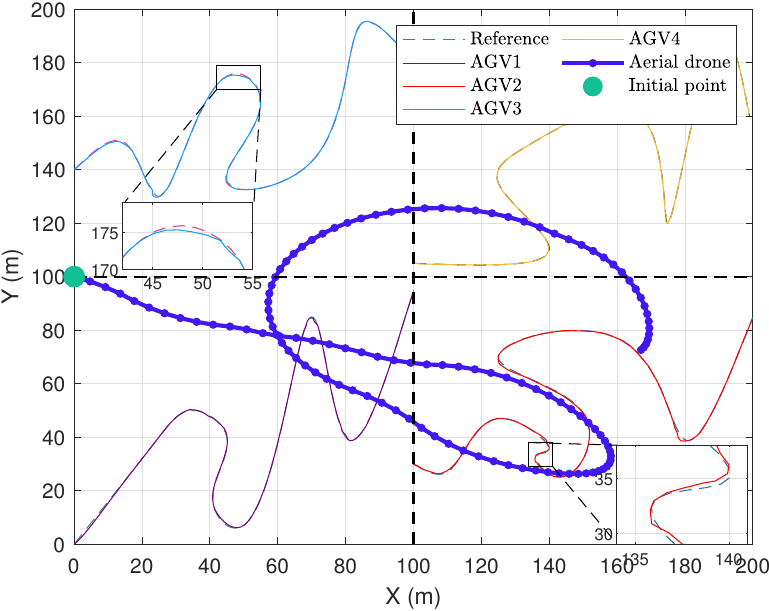}
        \caption{\label{UPb}Simulation with $4$ AGVs, EPA.} % 仅显示 (a)
    \end{subfigure}
    \hfill
    \\ \vspace{0.5cm}
        \begin{subfigure}{0.32\linewidth}
        \centering
        \includegraphics[width=\linewidth]{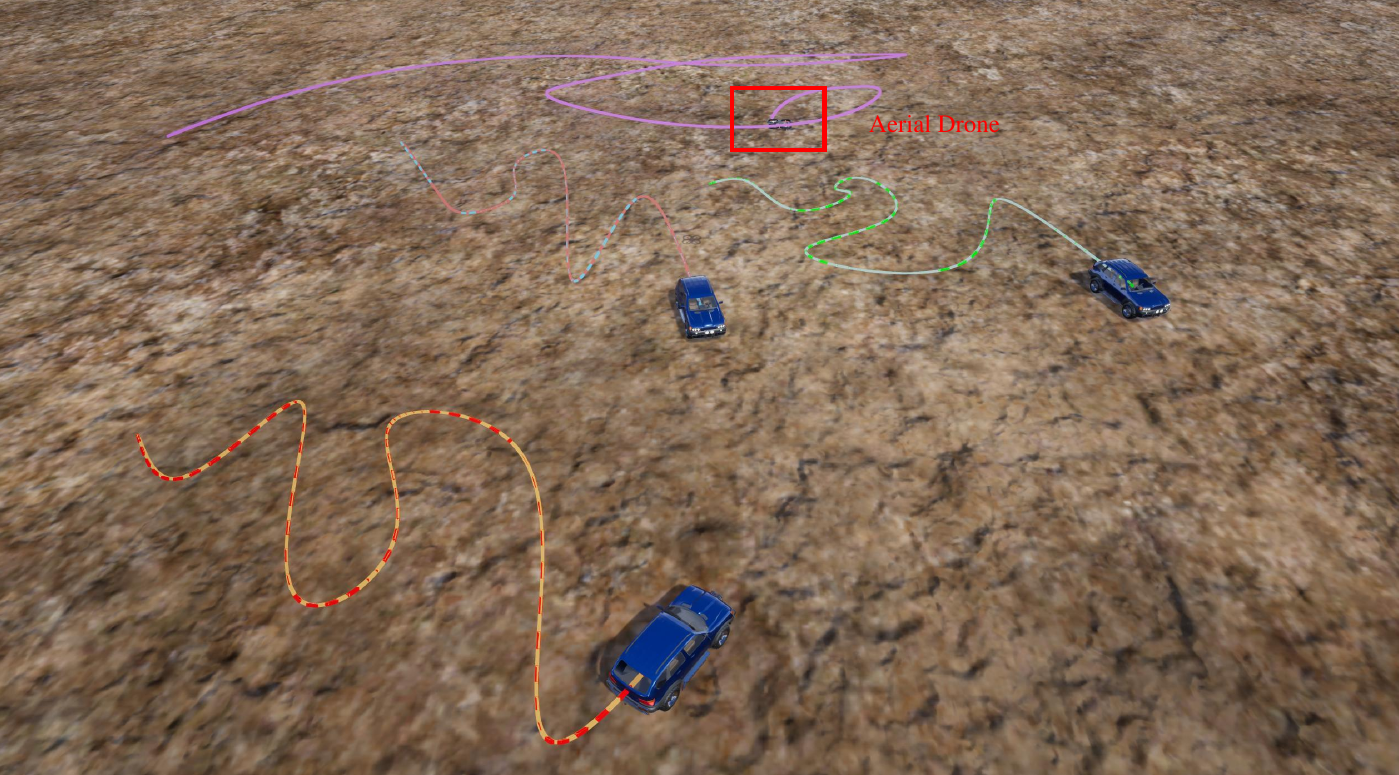}
        \caption{\label{UPcc}Validation with $3$ AGVs, Proposed.} % 仅显示 (a)
    \end{subfigure}
    \hfill
        \begin{subfigure}{0.32\linewidth}
        \centering
        \includegraphics[width=\linewidth]{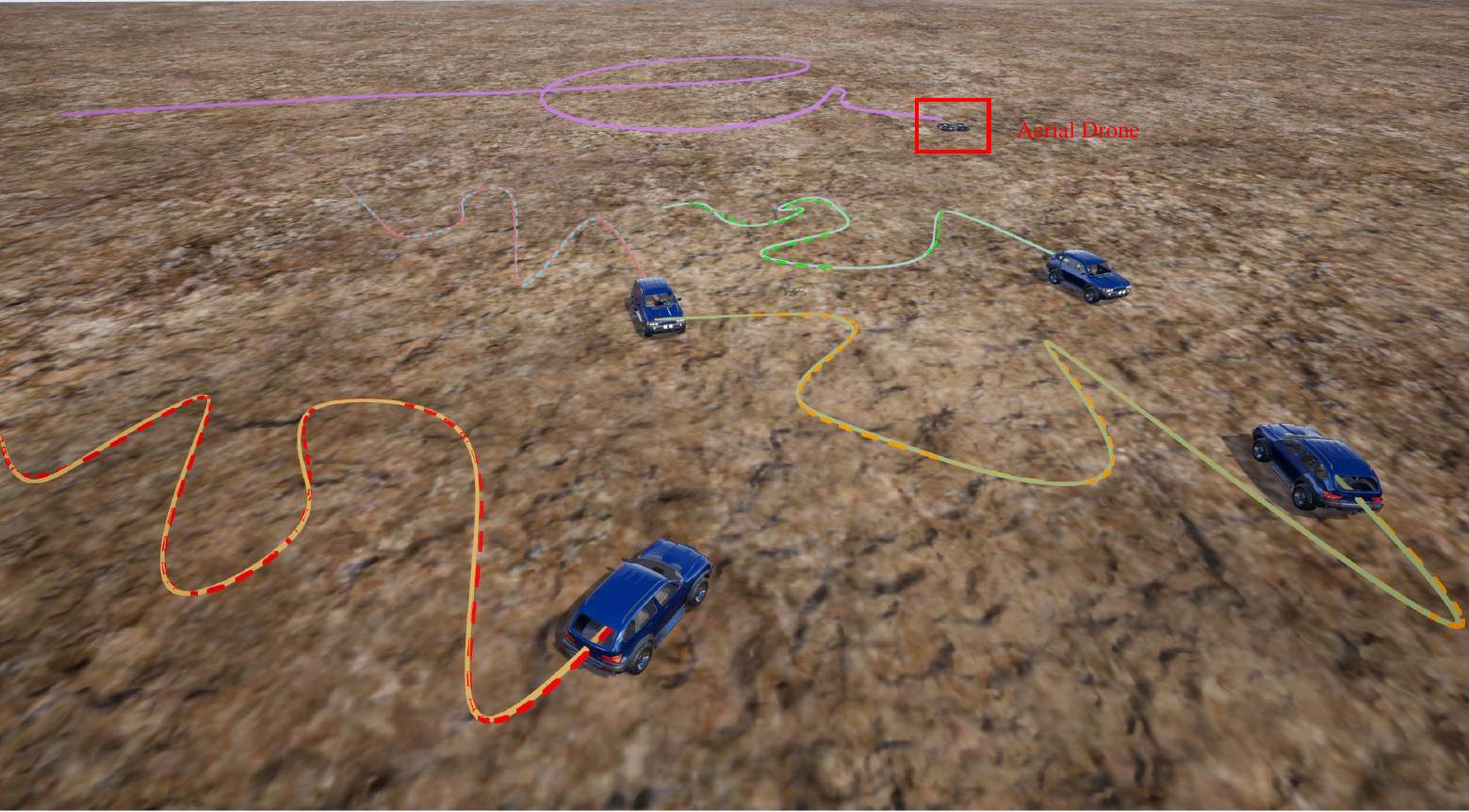}
        \caption{ \label{UPaa}Validation with $4$ AGVs, Proposed.} % 仅显示 (a)
    \end{subfigure}
    \hfill
    \begin{subfigure}{0.32\linewidth}
        \centering
        \includegraphics[width=\linewidth]{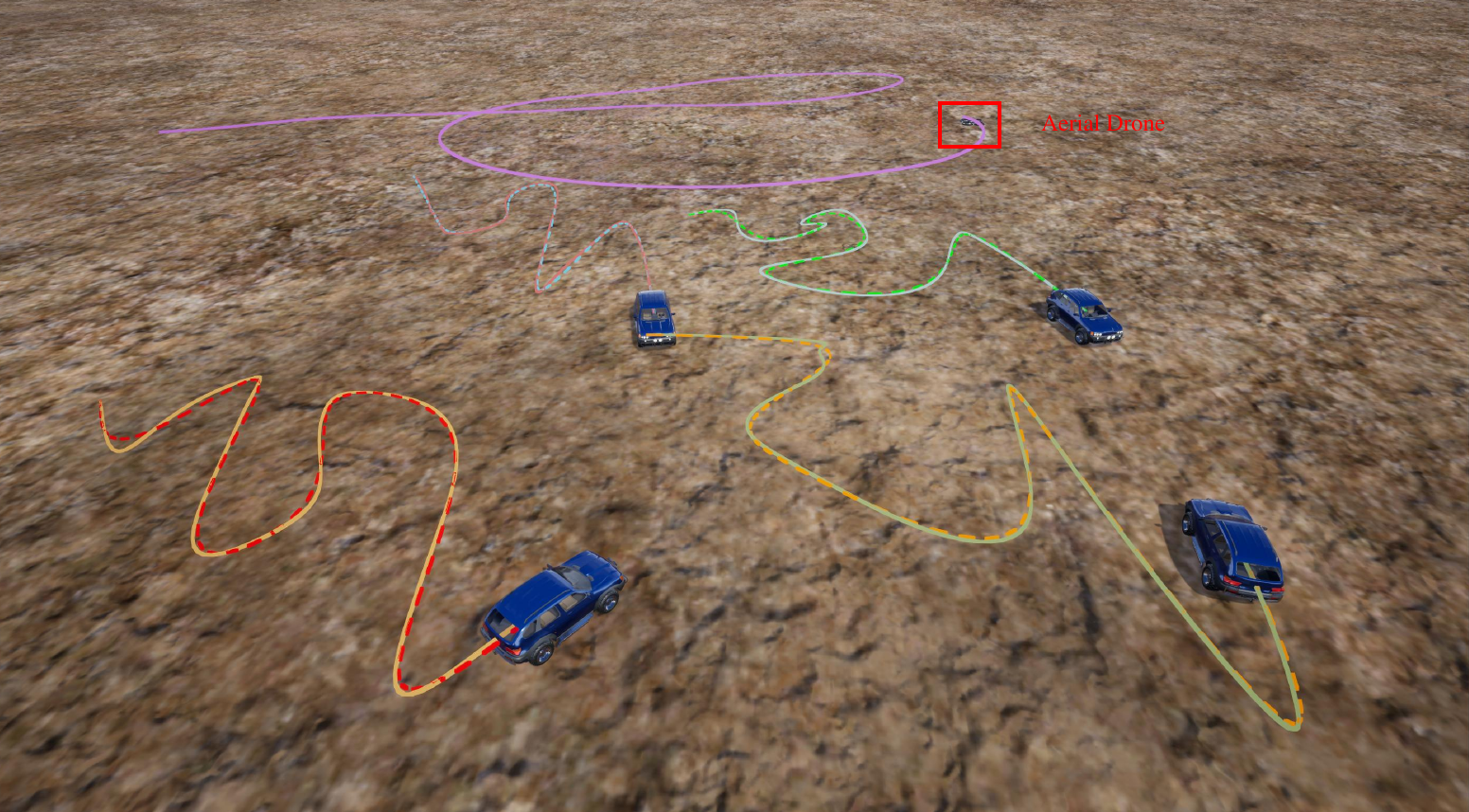}
        \caption{\label{UPbb}Validation with $4$ AGVs, EPA.} % 仅显示 (a)
    \end{subfigure}
    \hfill
\caption{\label{fig: Tracking path} The simulation and AirSim-based validation of drone trajectory and AGVs tracking paths under different scenarios.}
\end{figure*}

Fig. \ref{fig: Tracking path} compares the drone trajectories and AGV tracking results under three different scenarios. Fig. \ref{UPc} presents the result of the proposed method when only three AGVs are involved. In this case, the overall control task is relatively simple, allowing the drone to maintain a shorter trajectory while providing sufficient wireless support. As a result, the AGVs exhibit high tracking accuracy throughout the entire path, even around sharp turns or directional changes, demonstrating that with reduced task load, the system can allocate resources more effectively and ensure precise trajectory following. Fig. \ref{UPa} corresponds to the proposed joint optimization framework with four AGVs. As the number of AGVs increases, the control complexity becomes higher, especially in regions where multiple AGVs perform sharp turns or frequent directional changes. The drone trajectory is skewed toward the lower region, where AGV2 and AGV3 follow more complex paths, reflecting the algorithm's ability to prioritize areas with greater control demands and dynamically allocate more communication resources where needed. Despite increased complexity, the tracking performance remains satisfactory, although slight deviations occur near the initial stage and in high-curvature regions. In contrast, Fig. \ref{UPb} shows the result under the EPA baseline with four AGVs. Without spatial prioritization, the drone follows a significantly longer trajectory and fails to focus on critical areas with high tracking requirements. Consequently, the tracking performance is notably worse, especially in sharp-turn regions where precise control is needed. The AGVs deviate considerably from their reference paths, indicating limited communication support under uniform power allocation.

\subsection{AirSim Platform-Based Experiments}

\begin{table}[H]
\centering
\caption{\label{tab:airsim}AirSim Simulation Parameters}
\begin{tabular}{ll|ll}
\toprule
\textbf{Parameter} & \textbf{Value} & \textbf{Parameter} & \textbf{Value} \\
\midrule
Rain & 20\% & Dust & None \\
Road Wetness & 30\% & Fog & 15\% \\
Snow & 20\% & Wind Speed x-axis & 2 m/s \\
Road Snow & None & Wind Speed y-axis & 2.5 m/s \\
Falling Leaves & None & Wind Speed z-axis & None \\
\bottomrule
\end{tabular}
\end{table}

% \begin{figure}[h]
% \centering
% \includegraphics[width=0.9\linewidth]{a_plot_reference/airsim.pdf}
% \caption{\label{fig: airsim}Validation of the proposed approach on the AirSim platform.}
% \end{figure}

To further validate the effectiveness of the proposed method, we implement the considered LAWN scenario within the AirSim simulation platform \cite{AirSim}, as illustrated in Fig. \ref{UPcc}-Fig. \ref{UPbb}. To better emulate real-world disturbances, we configured the AirSim environment with multiple weather parameters, as summarized in Table~\ref{tab:airsim}. 
\textcolor{black}{Specifically, we introduced moderate environmental dynamics by setting the rain and snow intensities to 20\%, road wetness to 30\%, and fog to 15\%, while disabling dust, road snow, and falling leaves. In addition, wind effects were introduced by configuring directional wind speeds to 2 m/s, 2.5 m/s, and 0 m/s along the X, Y, and Z axes, respectively.  In this setup, AGVs are deployed with predefined nonlinear trajectories featuring diverse motion characteristics, including sharp turns and variable speeds. The drone is tasked with providing wireless control support while dynamically adjusting its trajectory and power allocation according to the proposed joint optimization algorithm. It is worth noting that the optimization is implemented in MATLAB using the CVX toolbox, and the computation time per control update remains well within the sampling interval, confirming the practical computational efficiency of the proposed method \cite{AirSim}.}
Compared to simulation, the AirSim validation exhibits more noticeable tracking deviations, particularly around sharp turns and abrupt motion changes, mainly due to the incorporation of realistic physical constraints such as actuator delays, vehicle dynamics, and limited control frequency. Under these conditions, both the AGVs and the drone encounter greater challenges in maintaining high-precision trajectory following. Nevertheless, the overall trends remain consistent with the simulation results, where the drone still tends to hover around regions of high control complexity and adaptively supports AGVs with greater tracking demands, further demonstrating the robustness and practicality of the proposed method in more physically grounded environments. 
% In addition, the AirSim-based experiments demonstrate that the proposed framework can still function well under more detailed physical conditions. By considering factors such as motion delays and limited control frequency, the validation provides a closer approximation to real deployment scenarios than simplified simulations, which helps to better evaluate the adaptability and robustness of the method and shows that it can handle more practical challenges in control and coordination.

\section{Conclusion\label{sec:5}}
This paper has investigated a real-time wireless control framework for drone-assisted LAWNs, with a particular focus on delay-sensitive AGV trajectory tracking under FBL transmission. We have developed the communication-assisted predictive control system, in which we first derived a closed-form expression for the outage probability under FBL transmission, and then we incorporated it into the MPC cost function. Based on this, we have formulated the joint optimization problem to optimize control inputs, power allocation, and drone trajectory, subject to transmit power budgets and drone mobility constraints. We have transformed the inherently non-convex problem into a QP model and solved it via an AO framework, which leverages PGD and SCA techniques. Rigorous analysis was conducted to evaluate the convergence behavior and computational complexity of the proposed solution. Extensive simulation results demonstrate that the proposed algorithm consistently outperforms baseline methods in control performance. Furthermore, its effectiveness is validated through additional experiments conducted on the AirSim platform, confirming its practical applicability in LAWN-enabled IoT scenarios.

\bibliographystyle{IEEEtran}
\bibliography{a_plot_reference/reference}

\vspace{-0.5cm}
\begin{IEEEbiography}[{\includegraphics[width=1in,height=1.25in]{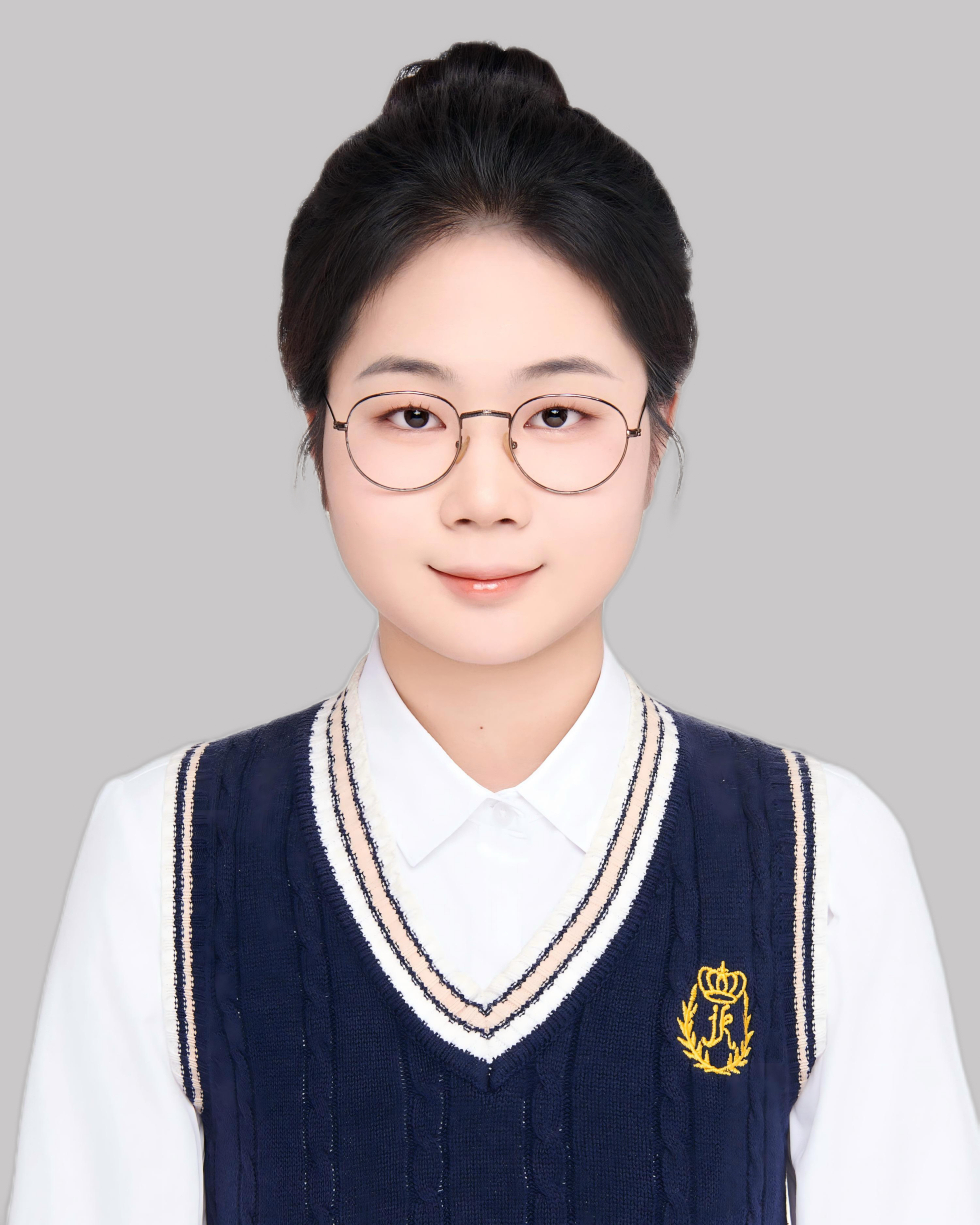}}]{Haijia Jin} received the B.Eng. degree in communication engineering from Nanjing University of Information Science and Technology, Nanjing, China, in 2024. She is currently pursuing the M.S. degree in the School of Automation and Intelligent  Manufacturing, Southern University of Science and Technology, Shenzhen, China. Her current research interests include low-altitude wireless networks, finite blocklength transmission, and optimization theory.
\end{IEEEbiography}

\vspace{-0.5cm}
\begin{IEEEbiography}
[{\includegraphics[width=1in,height=1.25in]{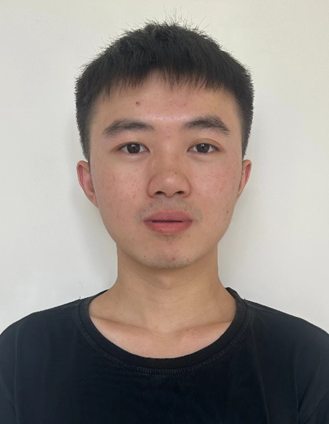}}]{Jun Wu} received the B.E. degree in the School of Electronic Engineering from  Southwest Jiaotong University, China, in 2021. He is currently working toward the Ph.D. degree in the School of Automation and Intelligent  Manufacturing,  Southern University of Science and Technology, Shenzhen, China. His research focuses on the area of integrated sensing and communications, UAV communications, low-altitude wireless networks, and convex optimization.
\end{IEEEbiography}

\vspace{-0.5cm}
\begin{IEEEbiography}[{\includegraphics[width=1in,height=1.25in]{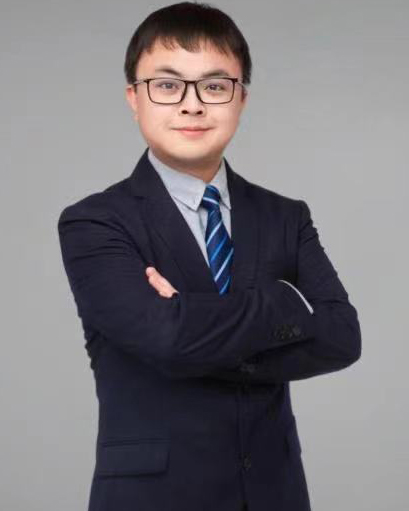}}]{Weijie Yuan} (Senior Member, IEEE) Weijie Yuan (Senior Member, IEEE) is now an Assistant Professor with the Southern University of Science and Technology. His research interests include Integrated Sensing and Communications (ISAC), Orthogonal Time Frequency Space (OTFS), and the Low-Altitude Wireless Networks (LAWN). He currently serves as an Editor for the IEEE Transactions on Wireless Communications, IEEE Transactions on Mobile Computing, IEEE Communications Magazine, IEEE Communications Standards Magazine, IEEE Transactions on Green Communications and Networking, IEEE Communications Letters, and IEEE Open Journal of Communications Society, an Guest Editor for IEEE Transactions on Vehicular Technology, IEEE Transactions on Network Science and Engineering, and IEEE Internet of Things Journal. He was the Track-Chair for IEEE ICC 2025 and IEEE VTC 2025-Spring. He served as an Organizer/the Chair of several workshops and special sessions in flagship IEEE and ACM conferences, including IEEE ICC, IEEE VTC, IEEE GlobeCom, IEEE/CIC ICCC, IEEE SPAWC, IEEE WCNC, IEEE ICASSP, and ACM MobiCom. He is the Founding Chair of the IEEE ComSoc Special Interest Group (SIG) on LAWN as well as the SIG on OTFS. He was listed among the World's Top 2\% Scientists by Stanford University for citation impact from 2021 to 2024, and among the Elsevier Highly-Cited Chinese Researchers. He was a recipient of the Best Editor from IEEE CommL, the Best Paper Award from IEEE ICC 2023, IEEE/CIC ICCC 2023, and IEEE GlobeCom 2024, as well as the 2025 IEEE Communications Society \& Information Theory Society Joint Paper Award. 
\end{IEEEbiography}

\begin{IEEEbiography}
[{\includegraphics[width=1in,height=1.25in]{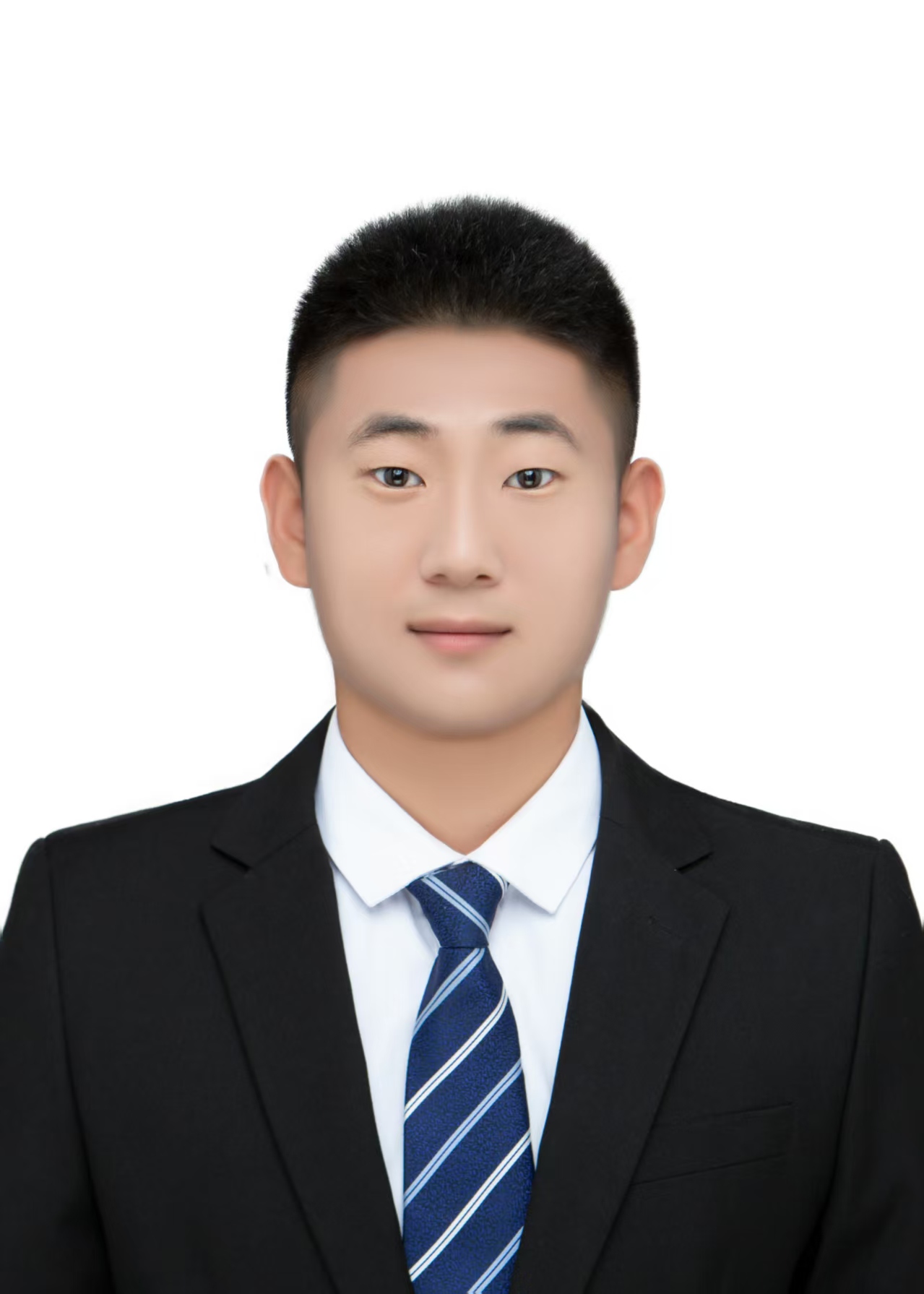}}]{Ruizhi Ruan} received the B.S. degree from the College of Information and Communication, Guilin University of Electronic Technology University, Guilin, China, in 2024. He is currently pursuing the master’s degree with the College of Semiconductors (National Graduate College for Engineers), Southern University of Science and Technology, Shenzhen, China. His research focuses on UAV motion planning, visual language navigation, and multi-agent exploration.
\end{IEEEbiography}

\begin{IEEEbiography}
[{\includegraphics[width=1in,height=1.25in]{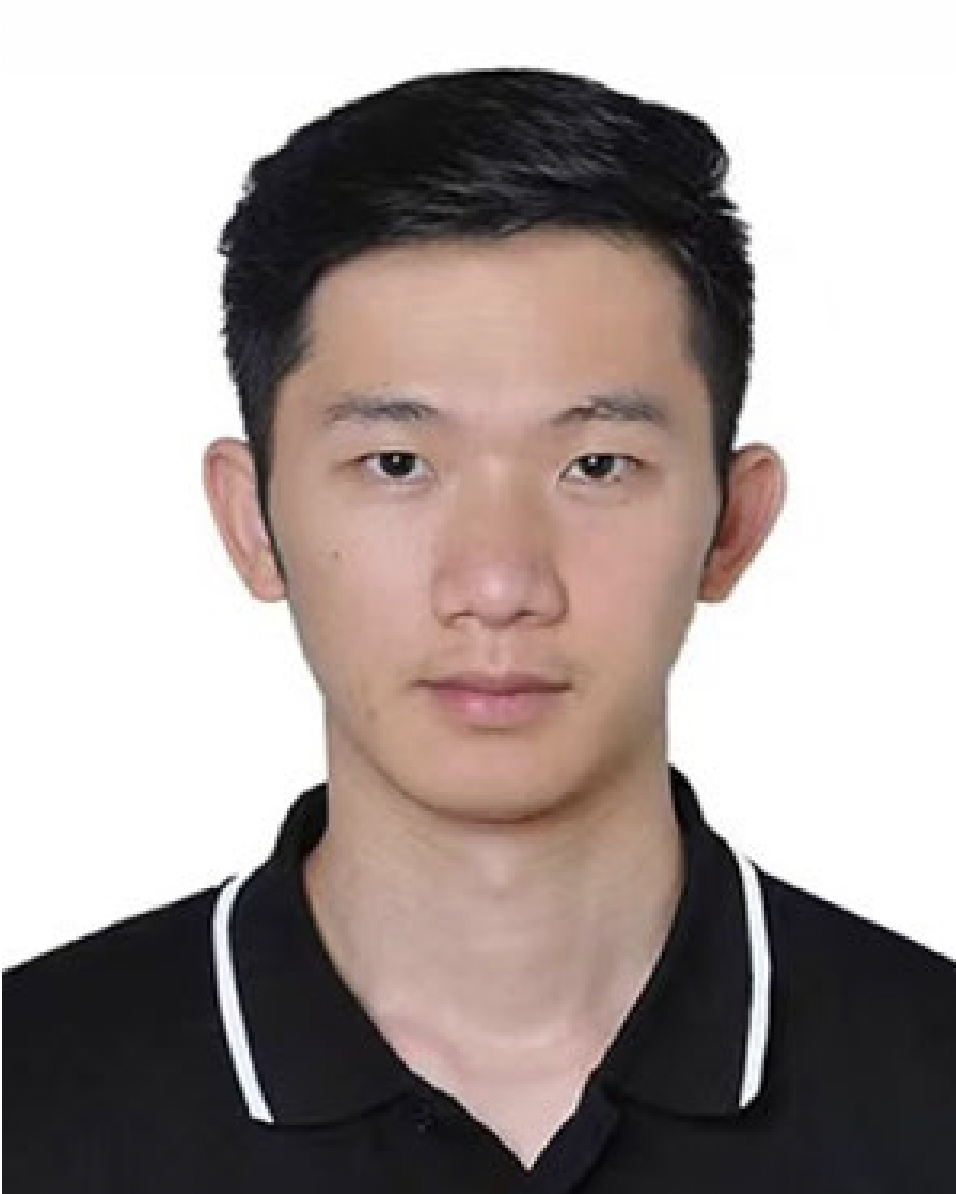}}]{Jiacheng Wang} (Member, IEEE) received the M.S.
and Ph.D. degrees from the School of Communication and Information Engineering, Chongqing University of Posts and Telecommunications, in 2018 and 2022, respectively. From 2021 to 2022, he was a Visiting Researcher with the College of Computing and Data Science, Nanyang Technological University (NTU), Singapore. He is currently with NTU. He has published more than 40 papers, including IEEE JOURNAL ON SELECTED AREAS IN COMMUNICATIONS, IEEE TRANSACTIONS ON MOBILE COMPUTING, IEEE TRANSACTIONS ON WIRELESS COMMUNICATIONS, IEEE TRANSACTIONS ON COGNITIVE COMMUNICATIONS AND NETWORKING, IEEE TRANSACTIONS ON VEHICULAR TECHNOLOGY, IEEE COMMUNICATIONS IN MATHEMATICS, OPTIMIZATION AND SYSTEMS THEORY, \emph{IEEE Wireless Communications Magazine}, IEEE Network Magazine, IEEE WIRELESS COMMUNICATIONS LETTERS, IEEE GLOBECOM, IEEE ICC, and IEEE WCNC. His research interests include generative AI, integrated sensing and communications, network optimization, and edge intelligence. He has received the IEEE ICC 2025 Best Paper Award. He was a Guest Editor of IEEE TRANSACTIONS ON COGNITIVE COMMUNICATIONS AND NETWORKING, IEEE TRANSACTIONS ON NETWORK SCIENCE AND ENGINEERING, \emph{Wireless Communications}, IEEE OPEN JOURNAL OF THE COMMUNICATIONS SOCIETY, \emph{IEEE Internet of Things Magazine}, and IEEE NETWORKING LETTERS. He serves as an Associate Editor for IEEE TRANSACTIONS ON NETWORK AND SERVICE MANAGEMENT and IEEE OPEN JOURNAL OF THE COMMUNICATIONS SOCIETY.
\end{IEEEbiography}

\begin{IEEEbiography}
[{\includegraphics[width=1in,height=1.25in]{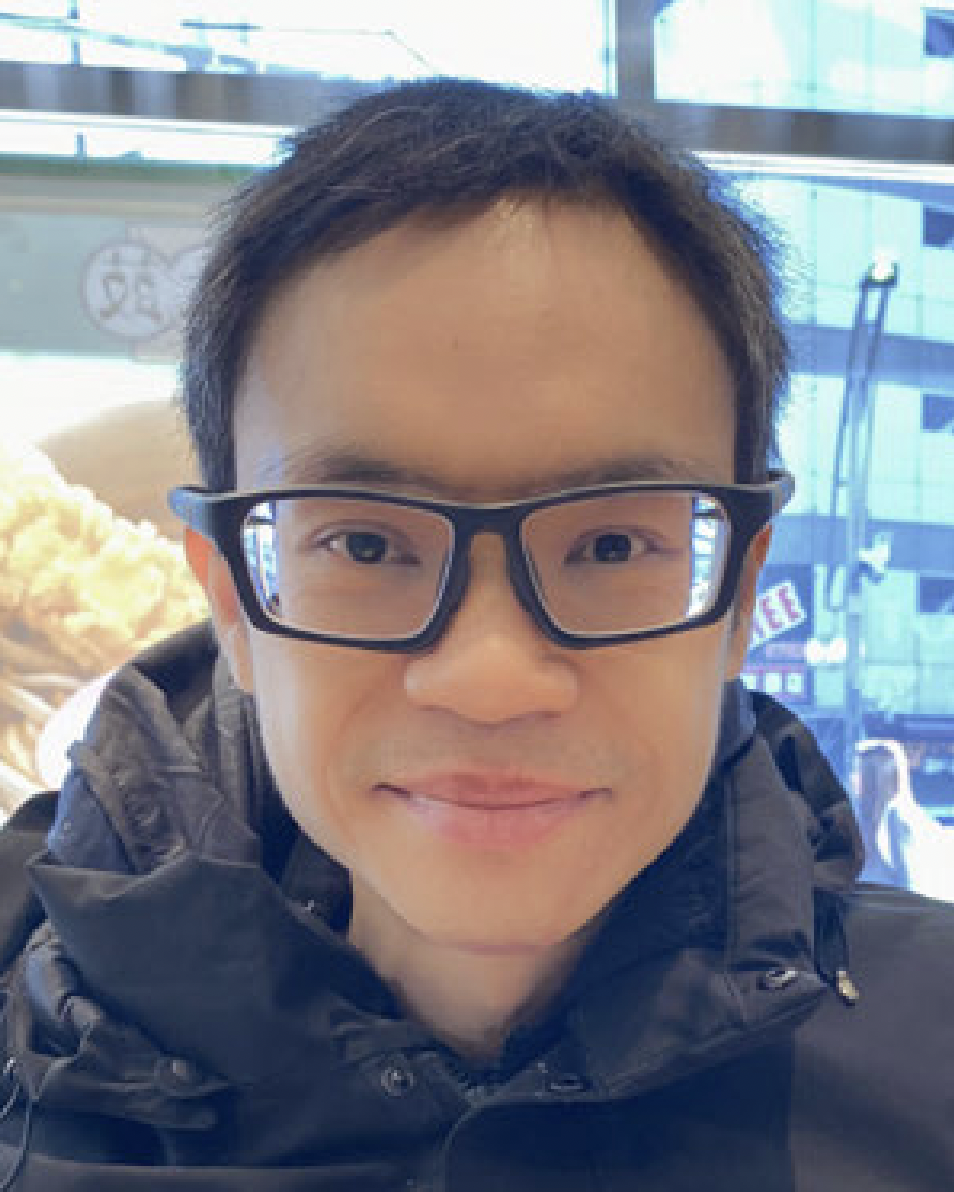}}]{Dusit Niyato} (Fellow, IEEE) received the B.Eng. degree from the King Mongkuts Institute of Technology Ladkrabang (KMITL), Thailand, and the Ph.D. degree in electrical and computer engineering from the University of Manitoba, Canada. He is currently a Professor with the College of Computing and Data Science, Nanyang Technological University, Singapore. His research interests include mobile generative AI, edge general intelligence, quantum computing and networking, and incentive mechanism design.
\end{IEEEbiography}

\begin{IEEEbiography}
[{\includegraphics[width=1in,height=1.25in]{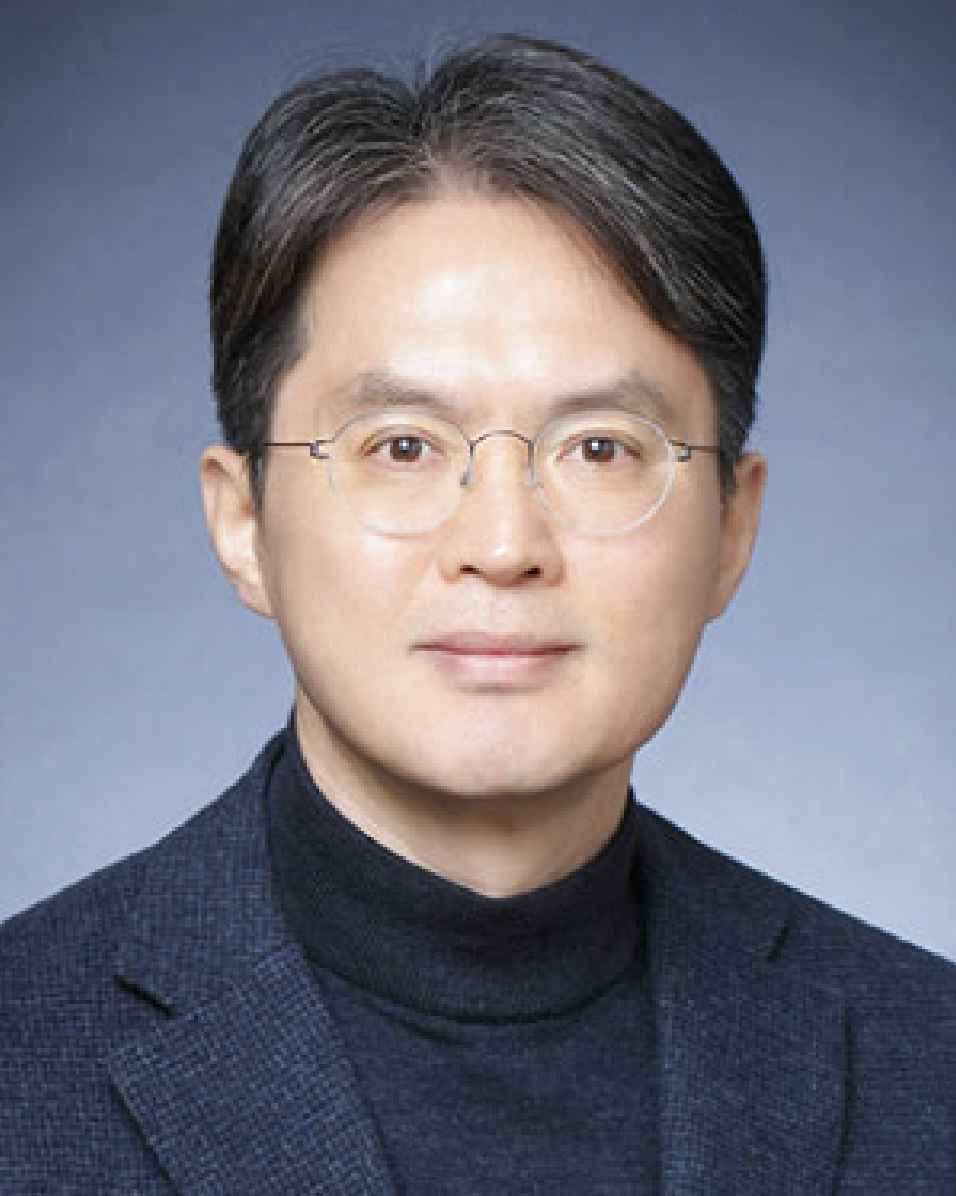}}]{Dong In Kim} (Life Fellow, IEEE) received the Ph.D. degree in electrical engineering from the University of Southern California, Los Angeles, CA, USA, in 1990. He was a tenured Professor with the School of Engineering Science, Simon Fraser University, Burnaby, BC, Canada. He is currently a Distinguished Professor with the College of Information and Communication Engineering, Sungkyunkwan University, Suwon, South Korea. He is a Fellow of Korean Academy of Science and Technology and a Life Member of the National Academy of Engineering of Korea. He was a recipient of the NRF of Korea Engineering Research Center (ERC) in Wireless Communications for RF Energy Harvesting from 2014 to 2021. He received several research awards, including the 2023 IEEE ComSoc Best Survey Paper Award and the 2022 IEEE Best Land Transportation Paper Award. He was selected as the 2019 recipient of the IEEE ComSoc Joseph LoCicero Award for Exemplary Service to Publications. He was the General Chair of the IEEE ICC 2022, Seoul. From 2001 to 2024, he served as an Editor, an Editor at Large, and an Area Editor of Wireless Communications I for IEEE TRANSACTIONS ON COMMUNICATIONS. From 2002 to 2011, he served as an Editor and a Founding Area Editor of Cross-Layer Design and Optimization for IEEE TRANSACTIONS ON WIRELESS COMMUNICATIONS. From 2008 to 2011, he served as the Co-Editor-in-Chief for IEEE/KICS JOURNAL OF COMMUNICATIONS AND NETWORKS. He served as the Founding Editor-in-Chief for IEEE WIRELESS COMMUNICATIONS LETTERS from 2012 to 2015. He has been listed as the 2020, 2022, and 2025 Highly Cited Researcher (HCR) by Clarivate Analytics.
\end{IEEEbiography}

\begin{IEEEbiography}
[{\includegraphics[width=1in,height=1.25in]{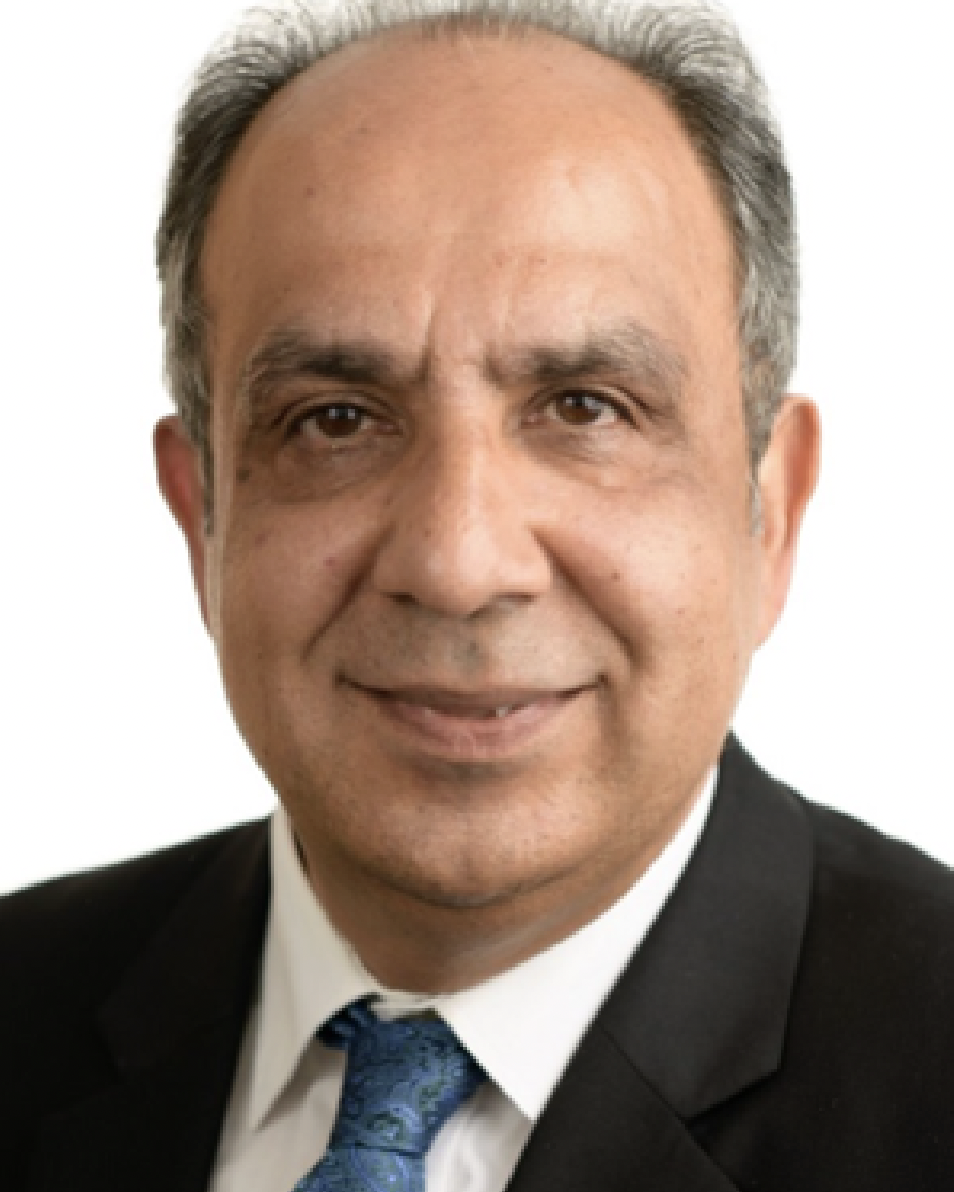}}]{Abbas Jamalipour} (Fellow, IEEE) received the Ph.D. degree in electrical engineering from Nagoya University, Nagoya, Japan, in 1996. He holds the position of a Professor of ubiquitous mobile networking with The University of Sydney and since January 2026, and the ComSoc Vice President for Membership and Global Activities. He has authored nine technical books, 11 book chapters, over 550 technical papers, and five patents, all in the area of wireless communications and networking. He is a fellow of the Institute of Electrical, Information, and Communication Engineers (IEICE) and the Institution of Engineers Australia, an ACM Professional Member, and an IEEE Distinguished Speaker. He was a recipient of several prestigious awards, such as the 2019 IEEE ComSoc Distinguished Technical Achievement Award in Green Communications, the 2016 IEEE ComSoc Distinguished Technical Achievement Award in Communications Switching and Routing, the 2010 IEEE ComSoc Harold Sobol Award, the 2006 IEEE ComSoc Best Tutorial Paper Award, and over 15 best paper awards. He was the President of the IEEE Vehicular Technology Society (2020–2021). Previously, he held the positions of the Executive Vice President and the Editor-in-Chief of VTS Mobile World and has been an Elected Member of the Board of Governors of the IEEE Vehicular Technology Society since 2014. He has been the General Chair and the Technical Program Chair of several prestigious conferences, including IEEE ICC, GLOBECOM, WCNC, and PIMRC. He was the Editor-in-Chief of IEEE TRANSACTIONS ON VEHICULAR TECHNOLOGY and IEEE WIRELESS COMMUNICATIONS, the vice president of conferences, and a member of the Board of Governors of the IEEE Communications Society. He sits on the editorial board of IEEE ACCESS and several other journals and is a member of Advisory Board of IEEE INTERNET OF THINGS JOURNAL.
\end{IEEEbiography}

\end{document}